\pdfoutput=1
\documentclass[manuscript]{rmaa}
\def\2F1{~_2F_1}
\usepackage{natbib}
\usepackage {graphicx} 
\title     
{
A turbulent model for  the surface 
brightness  of extragalactic jets
}
\author{L. Zaninetti\altaffilmark{1}}

\altaffiltext{1}{Dipartimento di Fisica Generale, Via Pietro Giuria 1,  
10125 Torino,  Italy (zaninetti@ph.unito.it.)}
\shortauthor{Zaninetti}
\shorttitle {Jet  Image}
\ReceivedDate{...............} 
\AcceptedDate{2} 
\SetYear{2000}
\resumen{Este documento describe ..
 } 

\abstract{
This paper  summarizes the known physics of turbulent jets 
observed in laboratory experiments. The formula,   which  gives the
power released in turbulence  describes the concentration of
turbulence/relativistic particles in each point of the
astrophysical jets. The same expression  is  also used to analyze 
the power released in turbulence in the case of pipe and non Newtonian
fluids. 
Through an integral operation 
it is possible
to deduce the intensity of synchrotron radiation for a profile
perpendicular  or not to  a straight jet , a 2D map for a
perpendicular , randomly oriented  straight jet as well as a 2D map
of complex trajectories such as NCC4061 and 3C31.
Presented here is a  simulation of the spectral index in brightness of 3C273 
as  well as a 2D map of the degree of linear polarization.
The Sobel  operator is applied to the theoretical
2D maps of straight perpendicular jets.
}
\addkeyword{galaxies :jets} 
\addkeyword{ radio continuum : galaxies} 
\def\apj{ApJ\,}
\def\apjl{ApJ\,}
\def\aap{A\&A\,}
\def\mnras{MNRAS\,}
\def\apjs{ApJS}
\def\jaa{J. Astrophys. Astr.}
\def\apss{Astrophysics and Space Science}

\begin{document}
\maketitle

\section{Introduction}

The  physical mechanism  that leads to the formation of the image 
in extra-galactic radio-sources can be the same   which  explains 
the physics of such objects ; this is  called   the unique model.
The unique model can be split  in two
\begin{enumerate}
\item 
The extremely relativistic, mono-energetic $e\pm$--pairs of high bulk
Lorentz factor, see~\citet{Kundt_1979,Kundt_2004,Kundt_2006}.
\item 
On modeling  the jets as intrinsically
symmetrical, relativistic, decelerating flows
it is possible  to simulate the radio-map of 3C31,
0326+39,1553+24 ,  NGC 315 and 3C296,
see \citet{Laing_2002,Canvin_Laing,Laing_2005,Laing_2006,Laing_2007}. 

\end{enumerate}

The transport processes are an open rather than a well established
field of research.
Following is a brief  review of some approaches.
\begin{enumerate}
\item
The  Kelvin-Helmholtz  instabilities   
  trigger a fluid 
turbulent cascade. 
The electrons are accelerated from the low wavelength 
of the turbulent spectrum , see~\citet{Zaninetti_1979,Eilek_2003}.
\item 
The laws of blob  containment , see \citet{Zaninetti_1989} ,
or the 3D random walk , see \citet{Zaninetti_1999} ,
produce simulated maps of  radiation intensity  
from extra-galactic radio-sources.

\item 
Plasma instabilities (e.g., Buneman, Weibel, and other two-stream
instabilities) created in collision-less shocks may be responsible 
for
particle (electron, positron) acceleration,
see~\citet{Nishikawa_2006}.

\end{enumerate}

The canonical approach to the turbulence is
through 
the  Navier-Stokes equations 
\begin{equation}
\frac {D  {\bf U}} {Dt} =
\frac {1}{\rho}\nabla p + \nu \nabla ^2 {\bf U} 
\quad ,
\end{equation}  
where ${\bf U}$ represents the velocity vector ,
$\frac {D  {\bf U}} {Dt} $ is the fluid particle acceleration ,
$ \nu$ is the kinematic viscosity,
$\rho$  the density 
and $p$ the pressure.
By applying
the Reynolds 
decomposition
 to the previous equations 
\begin{equation}
{\bf U(x,t)} = \langle {\bf U(x,t)}\rangle  + {\bf u(x,t)} 
\quad ,
\end{equation} 
where ${\bf u(x,t)} $ represents the fluctuation,
we obtain the equation of the mean momentum or Reynolds 
equations
\begin{equation}
\frac {{\overline D}  { U_j}} { \partial t}=
- \nu \nabla^2 (U_j)  
-\frac {\partial \langle u_i u_j \rangle} { \partial x_i} 
- \frac{1}{\rho} \frac {\partial \langle p \rangle  } { \partial x_i}
\quad ,
\end{equation}
where $\frac {{\overline D}} {\partial t}$ represents the mean
substantial derivative.
The difference between the Reynolds and the 
Navier-Stokes equations  is represented by the Reynolds stresses ,
$\frac {\partial \langle u_i u_j \rangle}{ \partial x_i} $.
The following  reviews three different models  which  adopt the 
turbulence in astrophysical jets.
\begin{enumerate}
\item 
A complex approach to the  turbulence 
is through  the  ensemble by~\citet{Favre1969} , 
see \citet{Bicknell1984} , where the density 
$\rho$ is expressed as  $\rho +\rho^{\prime}$ 
and the velocity $v_i$ as $v_i + v_i^{\prime}$
where ($i=1,2,3$).
From this double decomposition it is possible to obtain 
four equations for  mass , momentum , turbulent kinetic energy
and heat.
The behavior of the magnetic field is obtained by taking
into consideration the ensemble average of the MHD equations 
for the magnetic field $B_i$  in the infinite conductivity limit.
Once a law for the pressure of the atmosphere is considered,
it is possible
to obtain a law for the velocity , the electron distribution 
function , the magnetic field dependence and the density.
The above equations  allow us to deduce a law for the brightness 
along the jet $I_{\nu}^{\prime}$ once an adimensional 
opening angle , $ a^{\prime} = \phi / \phi_0 $ ,
is introduced.
The  brightness  law is of the type  $I_{\nu}^{\prime} \propto a^{-b}$
with $b ~[0.35-0.78]$ . 
In this model the opening angle  is variable , 
and the application
was done to $3C31$   which  shows a variable opening angle
  starting  from $12.3^{\circ}$ , reaching  a maximum
of $16.6^{\circ}$ in the middle 
and re-collimating  in the outer regions 
at $12.6^{\circ}$ , see Table~1 in~\citet{Bridle1980}.
Another application of this model 
was the study of  the physical state of the interstellar medium in
the optical counterpart NGC 383.
\item
A compressible k-e turbulence model was  used
by~\citet{Falle1994} to study 
the effect of turbulence on the propagation of
fluid jets.
In this model a connection between the opening angle of the 
jet  $\theta$ and the Mach number , $M$ , is 
\begin{equation}
M = \frac {1}{\theta} 
\quad .
\end{equation}
Through a numerical simulation  the z-component of velocity
along a cut perpendicular to the z-axis has been evaluated.
\item 
The near infrared images of HH~110 jet (Herbig Haro)
were interpreted as due to low velocity shocks produced by turbulent
processes, see ~\citet{Noriega-Crespo1996}.
The spatial intensity distribution of $H_2$ , $H_{\alpha}$ and
$[S_{II}]_{617/31}$ perpendicular to the flow axis and along 
the cross section of knots in $HH110$ has a behavior that 
can be approximated by a  Gaussian distribution , 
see Figure~4 in~\citet{Noriega-Crespo1996}.
In one case , $H_{\alpha}$ in knot P  of HH110, it 
is possible to see a bump near the maximum of the intensity
in the transversal direction. 
This can be considered the first observational evidence 
of a physical effect which  will  , later on , be called 
" valley on the top"  .
The results were explained by \citet{Noriega-Crespo1996}
adopting a turbulent mixing layer that followed  a Couette flow.
\end{enumerate}

The previous  approaches do  not solve the following questions
\begin {itemize}
\item
Can the physics of  turbulent jets , which  are observed
in  the laboratory,   be applied  to extra-galactic radio-sources~?
\item
Is it possible to build the image of a straight turbulent jet
which  is emitting synchrotron radiation from the law of dissipation
of  turbulence~?
\item 
Could the observed   cuts in  non-thermal radiative flux,
parametrized as a function of the distance ,  be 
compared with theoretical cuts~?  
\item
Could the 2D maps in brightness that characterize the complex 
morphologies associated with radio-galaxies be simulated~?  
\item 
Is it possible to simulate the variation of the 
spectral index in brightness of the non-thermal emission~?
\end{itemize}

The previous cited papers on turbulence solve some of the 
problems connected with the application of turbulence 
to astrophysical jets but leave other problems open
and  Table~\ref{problems} reports the status.
\begin{table}
 \caption {Synoptic table of the solved and unsolved 
           problems in three different turbulence papers and here }
 \label{problems}
 \[
 \begin{array}{lcccc}
 \hline
 \hline
 \noalign{\smallskip}
Problem  &  Bicknell ~1984   
& Falle~1994   
& Noriega-Crespo~et~al.~1996   & this~paper      \\
 \noalign{\smallskip}
 \hline
 \noalign{\smallskip}
brightness~  along~ the~ jet & yes & not & not & yes \\
perpendicular ~ brightness   & not & not & not & yes \\
image~as~integral~operation  & not & not & not & yes \\
law~for~magnetic~field       & yes & not & not & not \\
valley~on~the~top            & not & not & yes & yes \\
\noalign{\smallskip}
\noalign{\smallskip}
 \hline
 \hline
 \end{array}
 \]
 \end {table}

This paper briefly introduces in Section~\ref{intro_image}
an approximate law for the conversion of energy 
of the bulk flow into the turbulent cascade 
and the radiative transfer equation.
Section~\ref{fluids} computes the power released
in the turbulent cascade for three 
type of fluids :  turbulent fluid , pipe fluid
and {\it non-Newtonian fluid}.
Section~\ref{profiles}
computes through an integral,   the intensity profiles of the jet 
when it is directed in the direction perpendicular to the 
motion for the three types  of fluids previously 
analyzed.

Section~\ref{2D} deals with an algorithm used 
 to build the synchrotron image in two simple cases :
the straight jet is oriented in the direction perpendicular 
to the observer  or has a random direction.

Section~\ref{astro} presents  the images 
of the radio-jets which  
present bending and wiggling 
, such as   NGC4061 and 3C31.
This Section also 
contains a  simulation of  the  spectral index in brightness 
of turbulent astrophysical jets.

\section{Preliminaries to the   radio--images}
\label{intro_image}
\label{radio}
Section~\ref{efficiency} analyzes  a possible mechanism for 
 conversion of energy from
the flux of kinetic energy down to the turbulence.
 The matrix equation that implements
 the radiative transfer equation is introduced in
Section~\ref{radiative}~.

\subsection{The efficiency of energy conversion}

\label{efficiency}
The total  power , $Q$ , released  in a turbulent
cascade of the Kolmogorov type is, see~\citet{pelletier}
\begin{equation}
Q \approx \gamma_{KH} \rho  s_T^2 \quad ,
\end{equation}
 where
$\gamma_{KH} =  \gamma_{ad}  \frac{s_T}{a}$    
is the growth rate of K--H instabilities , $\rho$
is the matter density and $s_T$ is the sound velocity,
see Section~9.1 in~\citet{Zaninetti2007_b}.

The total maximum luminosity , $L_t$ , which  can be
 obtained for
the jet in a given region of radius $a_j$ and length $a_j$ is
\begin{equation}
L_t = \pi a_j^2 a_j Q = \gamma_{ad} \rho s_T^3 \pi a_j^2
\quad  .
\end{equation}
The mechanical luminosity  of  the jet  is
\begin{equation}
L_m  = \frac{1}{2}  \rho v^3 \pi a_j^2
\quad ,
\label {mechanical}
\end{equation}
and therefore the efficiency
of the conversion , $\chi_T$ ,
of the total available energy  in
turbulence is
\begin{equation}
\chi_T  = \frac {L_t}{L_m} = \gamma_{ad} \frac{2}{\pi} \frac
{1}{M^3} \approx \frac {1}{M^3} \quad ,
\end{equation}
where $M$ is the Mach number. From  this equation it is clear that
the fraction of the total available energy released firstly in the
turbulence and after in non thermal particles is a small fraction
of the bulk flow energy. 
The assumption made in
Section~3 and Section~4  by   \citet{Zaninetti2007_b}  in which
the bulk flow motion was treated independently from non--thermal
emission is now justified.

\subsection{Radiative transfer equation}

\label{radiative}
The transfer equation in the presence of emission only
, see for example~\citet{rybicki},
 is
 \begin{equation}
\frac { dI_{\nu}}{ds} =
j_{\nu}
\label{equazionetrasfer}
\quad ,
\end {equation}
where   $I_{\nu}$ is the specific intensity , $s$  is the line of
sight , $j_{\nu}$ the emission coefficient, and the index $\nu$
denotes the interested frequency of emission. 
The solution to 
equation~(\ref{equazionetrasfer}) is
\begin{equation}
 I_{\nu} (s) = I_{\nu} (s_0) +
\int_{s_0}^s   j_{\nu} (s\prime) ds\prime
\quad  ,
\end {equation}
where $I_{\nu} (s_0)$ represents the intensity  at
$s=s_0$ , in this  case zero.
In the case of synchrotron emission,
the maximum intensity of  mono-energetic
electrons in the
presence of a magnetic field $B$ in gauss is
\begin{equation}
 I_{\nu} (s) =
1.7 10^{-23}
 B_{\bot} \int C(s)ds ~~~
\frac {erg}{cm^2~s~sr~Hz}
\quad  ,
\end {equation}
where  $C(s)$  is the   number of electrons in a unit volume along
the line of sight and $B_{\bot}$ is the component of $B$ perpendicular
to the velocity vector , see for example equation~(5.550)
in~\citet{ginsburg}. In the case of  spatial dependence on 
$j_{\nu}$ it is  assumed
\begin{equation}
  j_{\nu}= C_j C(s)
\quad  ,
\end {equation}
where
\begin{equation}
C_j= 1.7 10^{-23}
 B_{\bot}  ~~~
\frac {erg}{~s~sr~Hz}
\quad  .
\end {equation}
The intensity in the non-homogeneous case is
\begin{equation}
 I_{\nu} (s) = C_j
\int_{s_0}^s   C (s\prime) ds\prime
\label{integral}
\quad  .
\end {equation}
The increase in brightness is proportional to the concentration
integrated along the line of  sight. In the Monte Carlo
experiments here analyzed,
 the concentration is memorized  on the
three-dimensional grid ${\mathcal C}$ and the intensity is
\begin{equation}
{\it I}(i,j) = \sum_k  \delta  \times
{\mathcal C}(i,j,k)
\quad , \label{transfer_sum}
\end{equation}
where $\delta$
is the spatial interval between
the  values of concentration  and  the sum is performed
over the interval of existence of  index k.

\section{Three types of fluids}  
\label{fluids}

 The cases of
shear layer produced by a turbulent jet, by a pipe 
and by a {\it non-Newtonian fluid} 
are now analyzed. 

\subsection{The turbulent jet}

\label{turbulent}
The theory of turbulent round jets  can be 
found in different textbooks.
The more important formulas are now reviewed 
as extracted 
from chapter~V by~\citet{Pope2000} ;
similar results can be found in~\citet{foot}
and in~\citet{Schlichting2004} .
We start with the centerline velocity $U_0(z)$ 
, equation~(5.6) in~\citet{Pope2000} ,  
as measured 
in the laboratory experiments :
\begin{equation}
\frac{U_0(z)} {U_1} = \frac {B}{z/d} 
\quad ,
\label{vlab}
\end{equation}
here $z$ denotes the main direction , 
$d$ is the diameter of the nozzle,
$B$ is a constant derived in the laboratory
that takes the value 5.8, and
$U_1$ is the initial jet  velocity.
The solution of the mean velocity $<U>$ 
, equation~(5.100) in~\citet{Pope2000} ,
along the main direction is
\begin{equation}
<U> =U_1 \frac { 8 a_T \nu_T}{z} \frac{1}{ (1 +a_T \eta^2)^2}
\label{uturb} 
\quad ,
\end{equation}
where  $\eta=\frac{r}{z}$ ,  $r$ is the radius of the jet at $z$,
$a_T$ is a constant  and 
$ \nu_T$ is the turbulent viscosity.
The viscosity
,
equation~(5.104) in~\citet{Pope2000}
,
 is
\begin{equation}
\nu_T = \frac{S}{8(\sqrt{2} -1)} 
\quad ,
\end{equation}
and  $a_T$, equation~(5.18)  in~\citet{Pope2000},
 is
\begin{equation}
a_T = \frac{(\sqrt{2} -1)} {S^2}
\quad ,
\end{equation}
where $S$ will be later defined.
The production of turbulent kinetic energy 
in the boundary layer approximation 
,
equation~(5.145) in~\citet{Pope2000} ,
is
\begin{equation}
\mathcal{P} = \nu_T (\frac{\partial <U>}{\partial y })^2
\quad ,
\label{pturb}
\end{equation}
where $y$ is a Cartesian coordinate that can be identified 
with $r$.
The flow rate of mass $m(z)$ is , see equation~(5.68) 
 in~\citet{Pope2000} ,
\begin{equation}
\dot {m(z)} = 2 \pi \rho b_{1/2} (b_{1/2}(z) U_0(z)) 
\times \int_{0}^{\infty} \xi f(\xi) d \xi
\quad ,
\label{mass}
\end{equation}
where 
\begin{equation} 
\xi = \frac{r} {b_{1/2}(z)}
\quad ,
\end{equation}
and 
\begin{equation}
f (\xi) = \frac{1}
{ 
(1 + A  \xi^2)^2
} 
\quad ,
\end{equation}
where $A$ is a constant that will be later defined
and $b_{1/2} $  is the value of the radius at which 
the velocity is half of the centerline value.
The jet draws  matter  from the surrounding mass of fluid.
Hence, the mass of fluid carried by the jet increases 
with the distance from the source. 
The previous formulas are exactly the same as in~\citet{Pope2000};
we now continue toward the astrophysical applications.
The quantity $S$ is connected with the opening angle 
$\alpha$ through the following relationship 
\begin{equation}
S = \tan \frac {\alpha}{2}
\quad .
\end{equation}
The self-similar solution for the velocity , 
equation~(\ref{uturb})  ,   can be re-expressed
introducing the half width $z=b_{1/2} /S $  
\begin{equation}
<U> =U_1 \frac { 8 a_T \nu_T}{z} \frac{1}{ (1 +A(\frac{r}{b_{1/2}})^2)^2} 
\quad ,
\end{equation}
where $A=\sqrt{2} -1 $~.
From the previous formula  the universal scaling of
the profile in velocity  is now clear.
From a careful inspection of the previous formula it is clear
that the variable $z$ should be expressed in $d$ units 
(the nozzle's diameter) in order to reproduce 
the laboratory results. In doing so we should 
find  the constant $k$ that allows us to deduce $B$ 
\begin{equation}
B = k \times   8 a_T \nu_T 
\quad .
\end{equation}

Table~\ref{data} reports a set of $S$ , $B$ and $\nu_T$ 
for different opening angles $\alpha$ .
\begin{table}
 \caption {Parameters of the turbulent jet
 when \lowercase{$k$}= 0.54 .  }
 \label{data}
 \[
 \begin{array}{ccccc}
 \hline
 \hline
 \noalign{\smallskip}
\alpha [rad] & \alpha [degree]   &  S  &  B & \nu_T      \\
 \noalign{\smallskip}
 \hline
 \noalign{\smallskip}
0.087 & 5     & 0.043 & 12.37 & 0.013  \\
0.185 & 10.64 & 0.093 &  5.79 & 0.028  \\
0.343 & 20    & 0.17  &  3.06 & 0.053  \\
0.523 & 30    & 0.26  &  2.01 & 0.08   \\
\noalign{\smallskip}
\noalign{\smallskip}
 \hline
 \hline
 \end{array}
 \]
 \end {table}
The assumption here used is that $k$ is the same for
different angles.
The velocity expressed in these practical units 
is 
\begin{equation}
<U> =\frac {B U_1 }{z}   
\frac{1}
{ 
(1 +A~(\frac{r}{b_{1/2}})^2)^2
} 
\label{uturbpract}
\quad .
\end{equation}

\begin{figure}
  \begin{center}
\includegraphics[width=10cm]{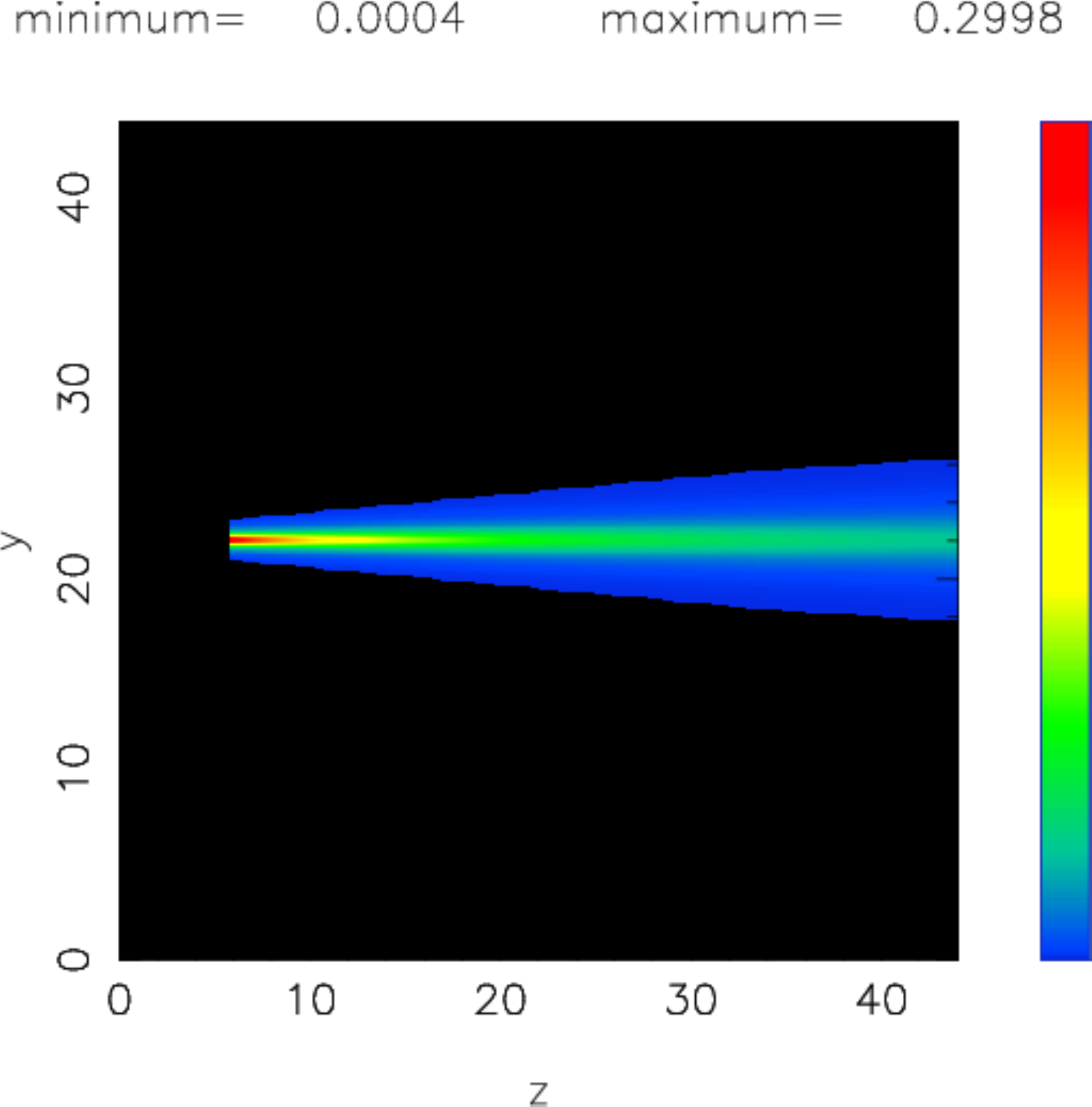}
  \end {center}
\caption {
Theoretical 2D map of the velocity , $<U> $ ,
given in units of the sound speed  when the  initial  Mach number is 0.3.
The length considered is $z/d=44$ , 
$k$ =0.54,
$A~$ = 0.414 
and  
$\alpha_{deg}=10.64$ .
          }%
    \label{utheo}
    \end{figure}

This formula can be used for $z$ expressed in $d$-units 
when  $z>B$ and Figure~\ref{utheo} reports 
the field of velocity for a laboratory jet.

The  first derivative 
of the profile in velocity as given by formula~(\ref{uturbpract})
 with  respect to the radius 
 is 
\begin{equation} 
\frac {d}{dr}  <U> = U_1 \times
\frac
{
-4\, \left( \sqrt {2}-1 \right) {\it A~ k}\,{{\it {b_{1/2}}}}^{6}A~r
}
{
\tan \left( 1/2\,\alpha \right)  \left( \sqrt {2}-1 \right) z \left( {
{\it {b_{1/2}}}}^{2}+A~{r}^{2} \right) ^{3}{{\it {b_{1/2}}}}^{2}
}
\quad .
\end{equation}

The production of turbulent kinetic energy is 
\begin{equation}
\mathcal{P} = \nu_T U_1^2 \times  
\frac
{
2\, \left( \sqrt {2}-1 \right) ^{2}{{\it A~ k}}^{2}{{\it {b_{1/2}}}}^{12}{A~
}^{2}{r}^{2}
}
{
\tan \left( 1/2\,\alpha \right)  \left( \sqrt {2}-1 \right) ^{3}{z}^{2
} \left( {{\it {b_{1/2}}}}^{2}+A~{r}^{2} \right) ^{6}{{\it {b_{1/2}}}}^{4}
}
\quad ,
\label{powerturb}
\end{equation}
and    $z$ should be expressed in $d$-units.
It is   interesting to note that the maximum of 
$ \mathcal{P} $  ,  is at 
\begin{equation}
r= \frac{1}{\sqrt{5}}  \frac{b_{1/2}} {\sqrt{A~}}
=0.69 {b_{1/2}}
\quad . 
\end{equation}

\begin{figure}
  \begin{center}
\includegraphics[width=10cm]{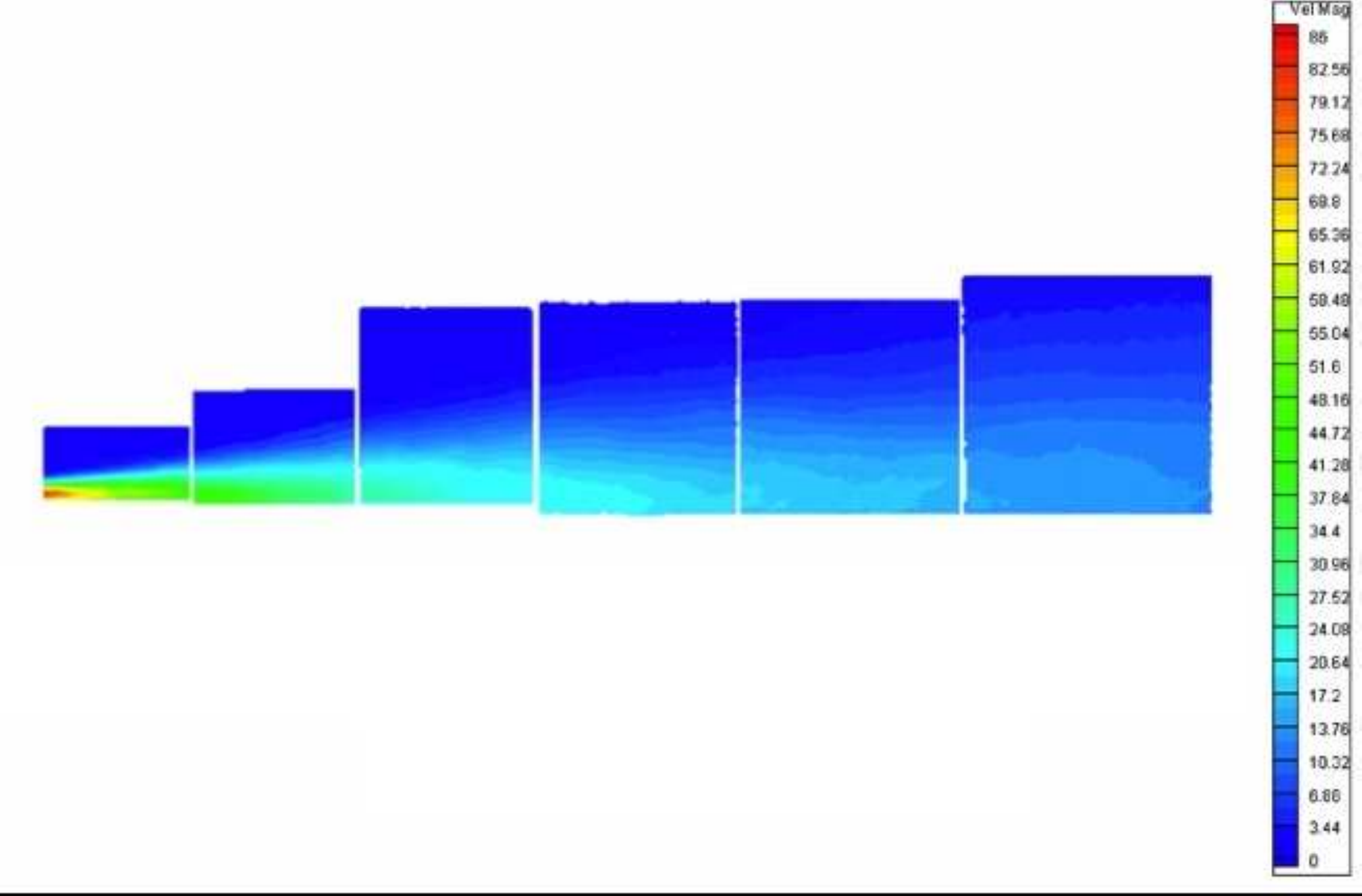}
   \end {center}
\caption {
Laboratory map of the turbulence intensity of Nitrogen jet,
Mach number=0.3.
Extracted from Figure~4.18 by~\citet{Bravo2006}.
          }%
    \label{tu}
    \end{figure}

\begin{figure}
  \begin{center}
\includegraphics[width=10cm]{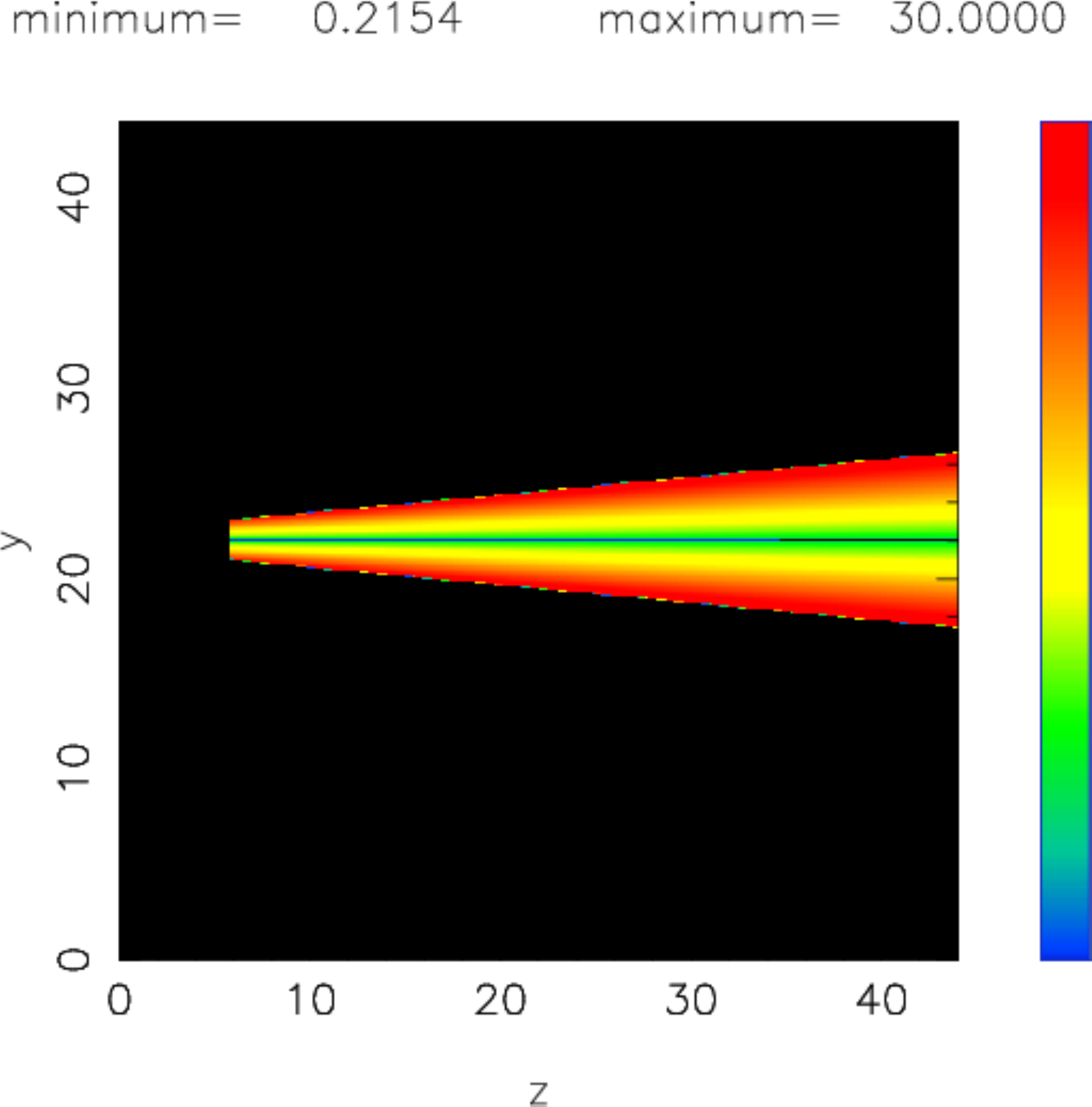}
  \end {center}
\caption {
Theoretical 2D map of the turbulence intensity , $Tu$ ,
when $Tu_{max}=30\%$.
The length considered is $z/d=44$ , 
$k$ =0.54,
$A~$ = 0.414 
$\rho$ =1 , 
and  
$\alpha_{deg}=10.64$ .
          }%
    \label{tutheo}
    \end{figure}

A quantity that is measured in the laboratory is the turbulence 
intensity $Tu$ 
\begin{equation}
Tu = \frac{U_{rms}}{<U>} 
\quad ,
\end {equation}
where $U_{rms}$ represents the standard deviation of the 
velocity measured in the laboratory and $<U>$
the averaged velocity.
This quantity is $\approx$  $25\%$ at $r=0$ 
and rises   with increasing r ,
see \citet{Pope2000} and \citet{Hussein1994}.
Recent results obtained using the PIV (Particle Image Velocimetry)
are able to produce  sophisticated measurements ,
see Figure~\ref{tu} , extracted from~\citet{Bravo2006} , who 
reports a  2D map of $Tu$ .
A possible theoretical explanation for  $Tu$ 
can be found from the only combination  
for $U_{rms}$   derived from ${\mathcal P}  $
which  produces an adimensional number when divided 
by ${<U>}$
\begin{equation}
Tu \sim  \frac
{ 
\sqrt{\frac{\nu_T}{\rho}   \frac{ \partial <U>}{\partial r}  }  
} 
{<U>}
\quad ,
\end{equation}
where $\rho$ is the density. 
Using the practical units 
\begin{equation}
Tu=
\frac 
{
\sqrt {2}\sqrt {{\frac { \left( \sqrt {2}-1 \right) {\it k}\,{{\it 
{b_{1/2}}}}^{4}A~r}{\rho\,z \left( {{\it {b_{1/2}}}}^{2}+A~{r}^{2} \right) ^{3}}}}
\tan \left( 1/2\,\alpha \right) z \left( {{\it {b_{1/2}}}}^{2}+A~{r}^{2}
 \right) ^{2}
}
{
2\, \left(  \sqrt {2}-1 \right) {\it k}\,{{\it {b_{1/2}}}}^{4}
}
\label{tu2d}
\quad .
\end{equation}
Figure~\ref{tutheo}
reports the theoretical 2D map of $Tu$   as
given by equation~(\ref{tu2d}) 
normalized 
to a maximum value $Tu_{max}$ as given  by  laboratory experiments.
When the flow rate of mass is expressed in these practical units
equation~(\ref{mass}) becomes  
\begin{equation}
\dot {m(z)} =
\frac{
\pi \,\rho\,z \left( \tan \left( 1/2\,\alpha \right)  \right) ^{3}{
\it U_1}\,d
}
{
\sqrt {2}-1
}
\quad .
\label{flow_practical}
\end{equation}
In this  formula    both $z$ 
and $d$
are present and
it is possible  to speak of physical units,
 but $z \ge d$ .

\subsection {Pipe fluids}

\begin{figure}
  \begin{center}
\includegraphics[width=10cm]{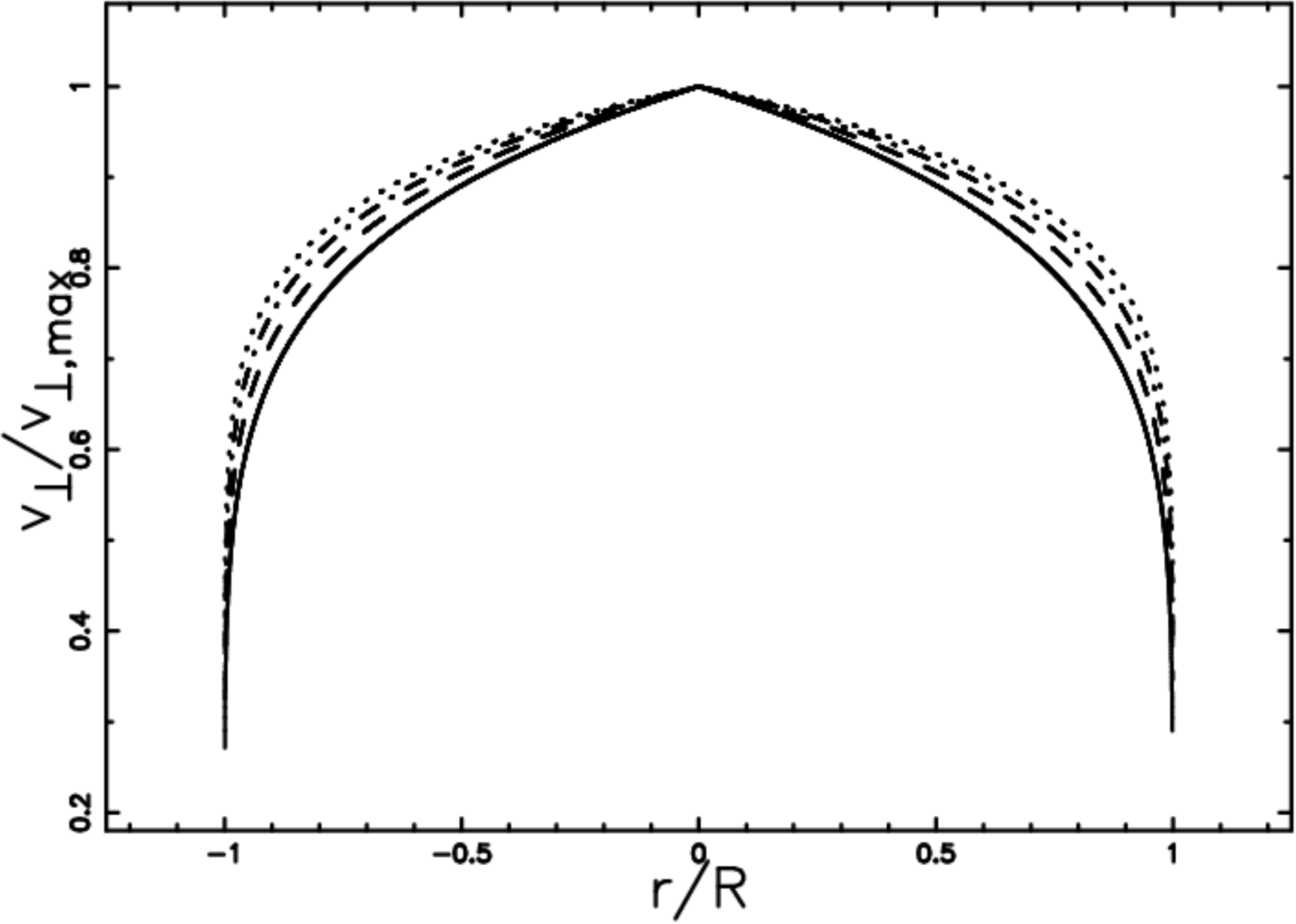}
  \end {center}
\caption {
Mean velocity profile ( in relative units ) vs. jet radius (in
relative units) when
n=6 ( full line ),
n=7 ( dashed ),
n=8 ( dot-dash-dot-dash ),
n=9 ( dotted ) and
n=10( dash-dot-dot-dot ).
          }%
    \label{prof_velocity}
    \end{figure}

In smooth circular tubes   the averaged
velocity for turbulent flow 
in the direction perpendicular
to the motion $v_{\bot}$  is often
parameterized through the following equation
\begin{equation}
\frac {\overline{v_{\bot}}} {v_{\bot,max}} =
 (1 - \frac {r}{a}) ^{\frac {1}{n}}
,
\label{eqnvz}
\end {equation}
where  $r$ is the distance from the jet  central axis ,
$a$  is the jet  radius,
$v_{\bot,max}$ is the maximum radial velocity  and $n$  is 
an integer,
see for example~\citet{foot}.
Figure~\ref{prof_velocity} reports 
the mean velocity profile
for different values of $n$.

\begin{figure}
  \begin{center}
\includegraphics[width=10cm]{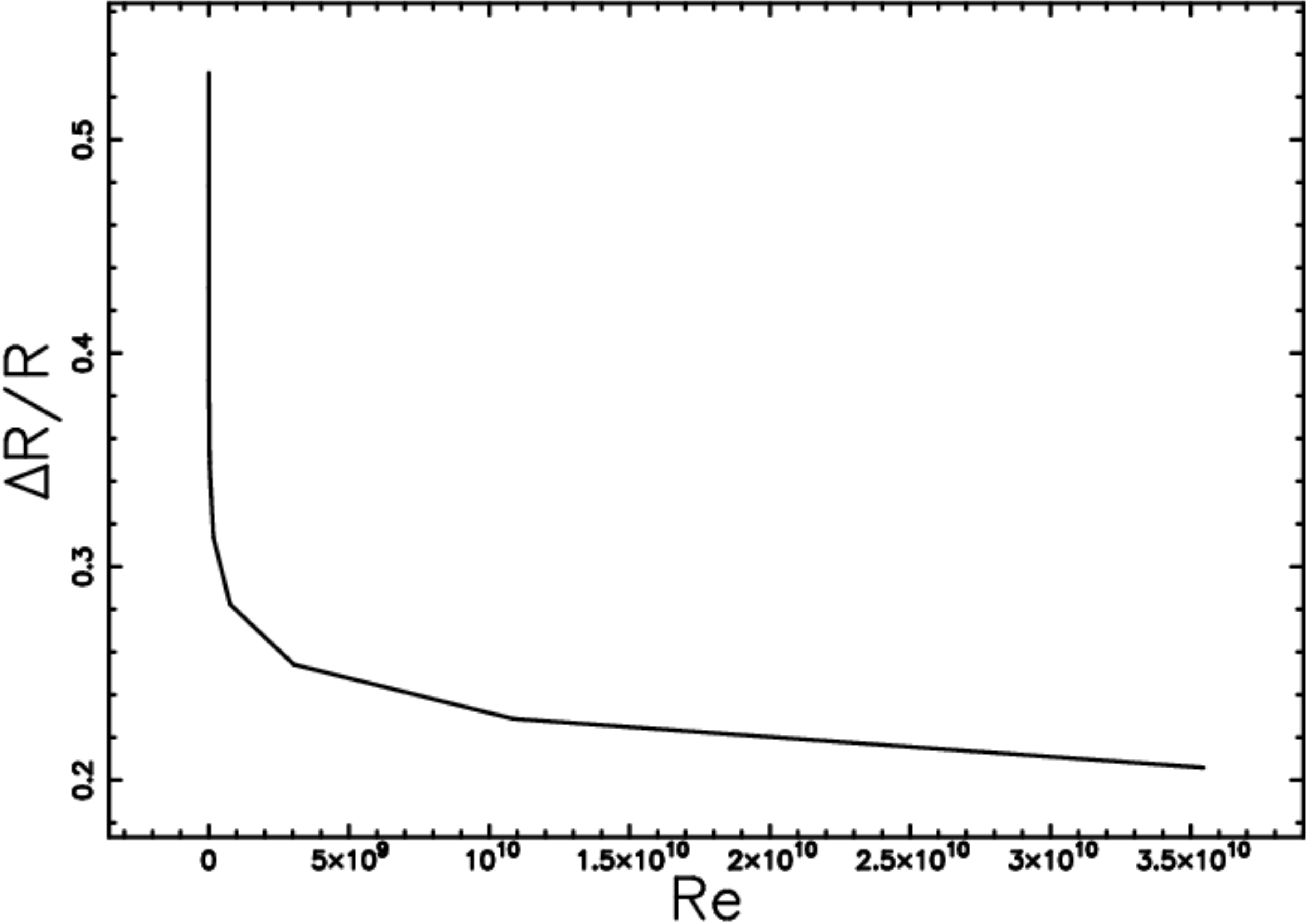}
  \end {center}
\caption {
$\frac {\Delta R}{R}$  vs.  Reynolds number when
$\frac {\overline{v_{\bot}}} {v_{\bot,max}}$=0.9 .
          }%
    \label{deltar}
    \end{figure}

On introducing $\Delta R=a-r$  , the following is found:
\begin{equation}
\frac {\Delta R}{R}=
(\frac {\overline{v_{\bot}}}
{v_{\bot,max}})^{\frac {1}{n}}
.
\label{form_deltar}
\end {equation}
Further on  $n$  is connected  to the Reynolds number ,$Re$,
through the following relationship
\begin{equation}
n = NINT ( 10^{a_{Re}} \times R^{b_{Re}})
,
\end {equation}
where NINT represents the nearest integer. The couple
($a_{Re},b_{Re}$)
is  easily found once the experimental correspondence
n $\longrightarrow Re$ is  provided , see for example
Table~\ref{equivalence} or ~\citet{foot}.
On applying the least square fitting procedure , 
$a_{Re}=$0.561 and   $b_{Re}=$0.0582.
Once a value  of
$\frac {\overline{v_{\bot}}} {v_{\bot,max}}$ is chosen   we can
plot  $\frac {\Delta R}{R}$  as a function of the Reynolds number,
see Figure~\ref{deltar}~.
Table~\ref{equivalence}   summarizes  the equivalence
between $n $, $Re$ and  $\frac {\Delta R}{R}$   when
$\frac {\overline{v_{\bot}}} {v_{\bot,max}}$ is fixed .
 \begin{table}
 \caption {Equivalence between parameters
 when  $\frac {\overline{v_{\bot}}} {v_{\bot,max}}$=0.9  .}
 \label{equivalence}
 \[
 \begin{array}{ccc}
 \hline
 \hline
 \noalign{\smallskip}
  n  &  Re  &  \frac {\Delta R}{R}   \\
 \noalign{\smallskip}
 \hline
 \noalign{\smallskip}
6 & 4   \times 10^3    & 0.53  \\
7 & 1.1 \times 10^5    & 0.47 \\
8 & 1.1 \times 10^6    & 0.43 \\
9 & 3.2 \times 10^6    & 0.38 \\

\noalign{\smallskip}
\noalign{\smallskip}
 \hline
 \hline
 \end{array}
 \]
 \end {table}
The first derivative of the profile 
of perpendicular  velocity as given by formula~(\ref{eqnvz})
is then computed 
\begin{equation} 
 \frac {1}{dr} \frac {\overline{v_{\bot}}} {v_{\bot,max}} = -
 \frac{1}{n} (1-\frac{r}{a})^{\frac{1}{n}} \frac {1}{a-r}
\quad  . 
\end{equation}
By analogy  with formula~(\ref{pturb}) 
 a power released in the cascade as 
\begin{equation}
\mathcal{P}_n = \nu_n (
\frac{\partial v_{\bot }}{\partial r})^2
\quad ,
\label{pn}
\end{equation}
is introduced 
where the index $n$ stands for {\it pipe}  fluid.
We therefore have 
\begin{equation}
\mathcal{P}_n = 
\frac
{
{\it \nu_n}\, \left({v_{\bot,max}}   \left( -{\frac {-a+r}{a}} \right) ^{{n}^{-1}}
 \right) ^{2}
}
{
{n}^{2} \left( -a+r \right) ^{2}
}
\quad .
\label{powern}
\end{equation}

\subsection{Non-Newtonian fluid}

In a {\it non-Newtonian fluid} the shear stress ,
$\tau_{yr}$ ,
 is  often
parameterized as 
\begin{equation}
\tau_{yr}=  k_n ( - \frac{du}{dr})^n 
\quad ,
\end{equation}
where $k_n$ is a constant , $u$ the velocity , $r$ the 
distance from the center  and $m$ an integer.
The profile in velocity results  to 
be
\begin{equation}
\frac{u(r)} {u_{max}} =
1 - (\frac{r}{a})^{\frac {m+1}{m}} 
\quad ,
\end {equation}
where $a$  is the 
radius of the pipe; 
see Problem 7-1 in \citet{transport1985} or Section~12.3 in 
\citet{hughes}.
Also here we   compute the first derivative of the profile 
of perpendicular  velocity 
\begin{equation}
\frac {d}{dr}
\frac{u(r)} {u_{max}} = -  
\frac {
\left( {\frac {r}{a}} \right) ^{{\frac {m+1}{m}}} \left( m+1 \right) 
 }
{ mr         }
\quad  .
\end{equation}
The power released in the cascade is 
\begin{equation}
\mathcal{P}_{non} = \nu_{non} (
\frac{\partial u(r) }{\partial r})^2
\quad ,
\label{pnon}
\end{equation}
where the index $non$ stands for {\it non-Newtonian }  fluid.
We therefore have 
\begin{equation}
\mathcal{P}_{non} = 
\frac
{
{\it  \nu_{non}}\, \left(u_{max}  \left( {\frac {r}{a}} \right) ^{{\frac {m+1}{m}}}
 \right) ^{2} \left( m+1 \right) ^{2}
}
{
{m}^{2}{r}^{2}
}
\quad .
\label{powernon}
\end{equation}

\section{Profiles in the perpendicular direction }

\label{profiles}
The following  explores how  the profile in 
intensity of a jet intersected by a plane perpendicular
to the direction of motion should appear.

Therefore, a classic Sobel filter is applied 
in order to find gradients in the profiles.

\begin{figure}
  \begin{center}
\includegraphics[width=10cm]{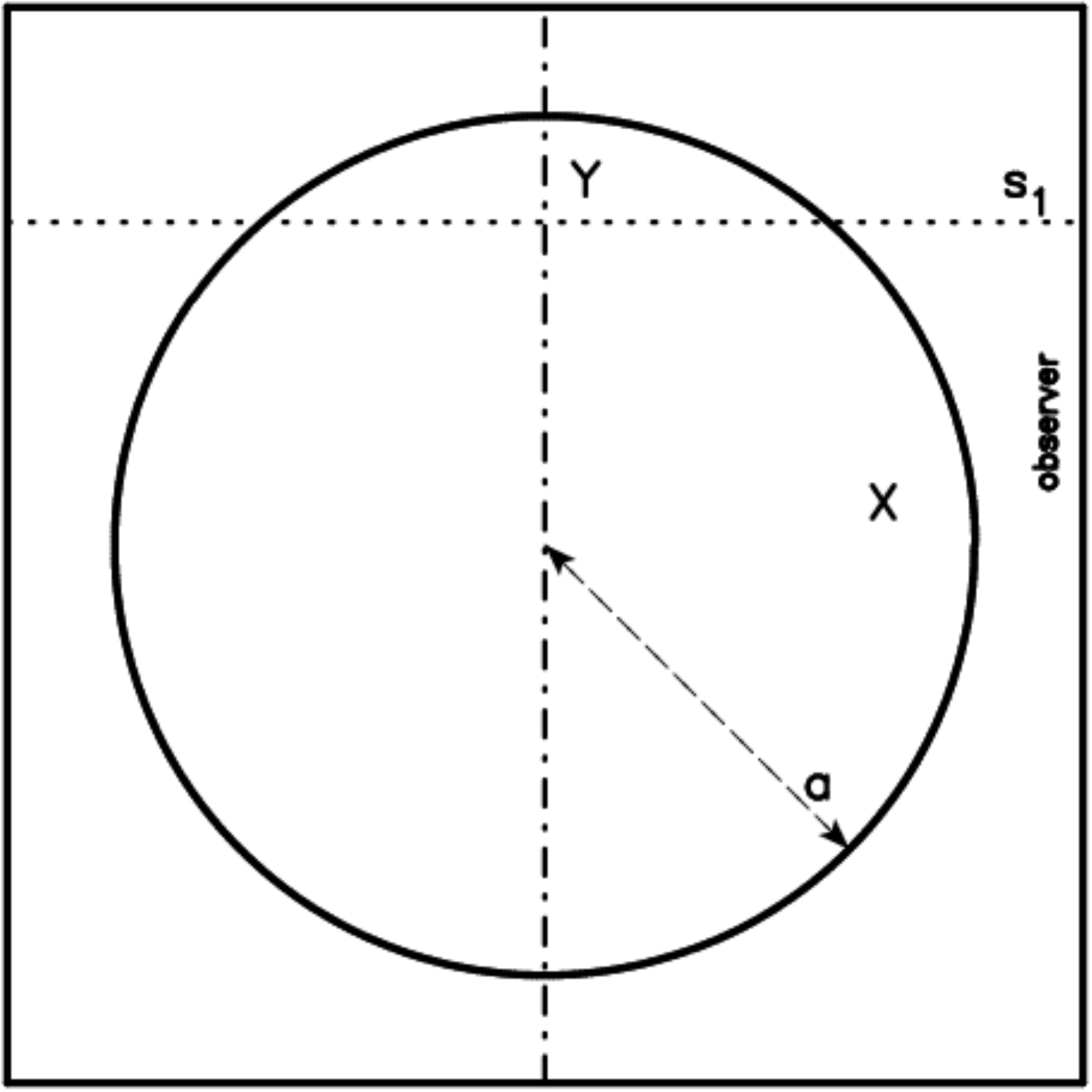}
  \end {center}
\caption {
The conical  source is  represented through
a section perpendicular to the jet  axis.
The observer is situated along the x direction,
one line  of sight is indicated.
          }%
    \label{asolo}
    \end{figure}

It is assumed   that the synchrotron-emissivity 
scales  as the power released in turbulent kinetic energy,
see equation~(\ref{powerturb}),
\begin{equation}
\epsilon  \sim  \mathcal{P}
\quad . 
\end{equation}
The cross section of the  jet is reported 
in Figure~\ref{asolo}.
Due to the additivity of the synchrotron emission 
along the line of sight, an integral 
operation is performed in order to obtain the intensity of emission
\begin{equation}
I(y)=
\int _{0}^{\sqrt {{a}^{2}-{y}^{2}}}
2 \times \epsilon (r)   {dx} 
\quad ,
\end{equation}
with  $r=\sqrt{x^2 +y^2}$ and $a$ representing the jet  radius.

\subsection{Profiles of the turbulent jet}

The intensity of emission  
according to formula~(\ref{powerturb}) is 
\begin{equation}
I(y)     \sim 
\int _{0}^{\sqrt {{{\it a}}^{2}-{y}^{2}}}\!4\,{\frac { \left( {{\it 
x}}^{2}+{y}^{2} \right) {A~}^{2}{{\it {b_{1/2}}}}^{8}{{\it a}}^{2}}{
 \left( {{\it {b_{1/2}}}}^{2}+A~{{\it x}}^{2}+A~{y}^{2} \right) ^{6}{z}^{2
} \left( \sqrt {2}-1 \right) \tan \left( 1/2\,\alpha \right) }}{d{\it 
x}}
\quad  .
\label{iyturb} 
\end{equation}

\begin{figure}
  \begin{center}
\includegraphics[width=10cm]{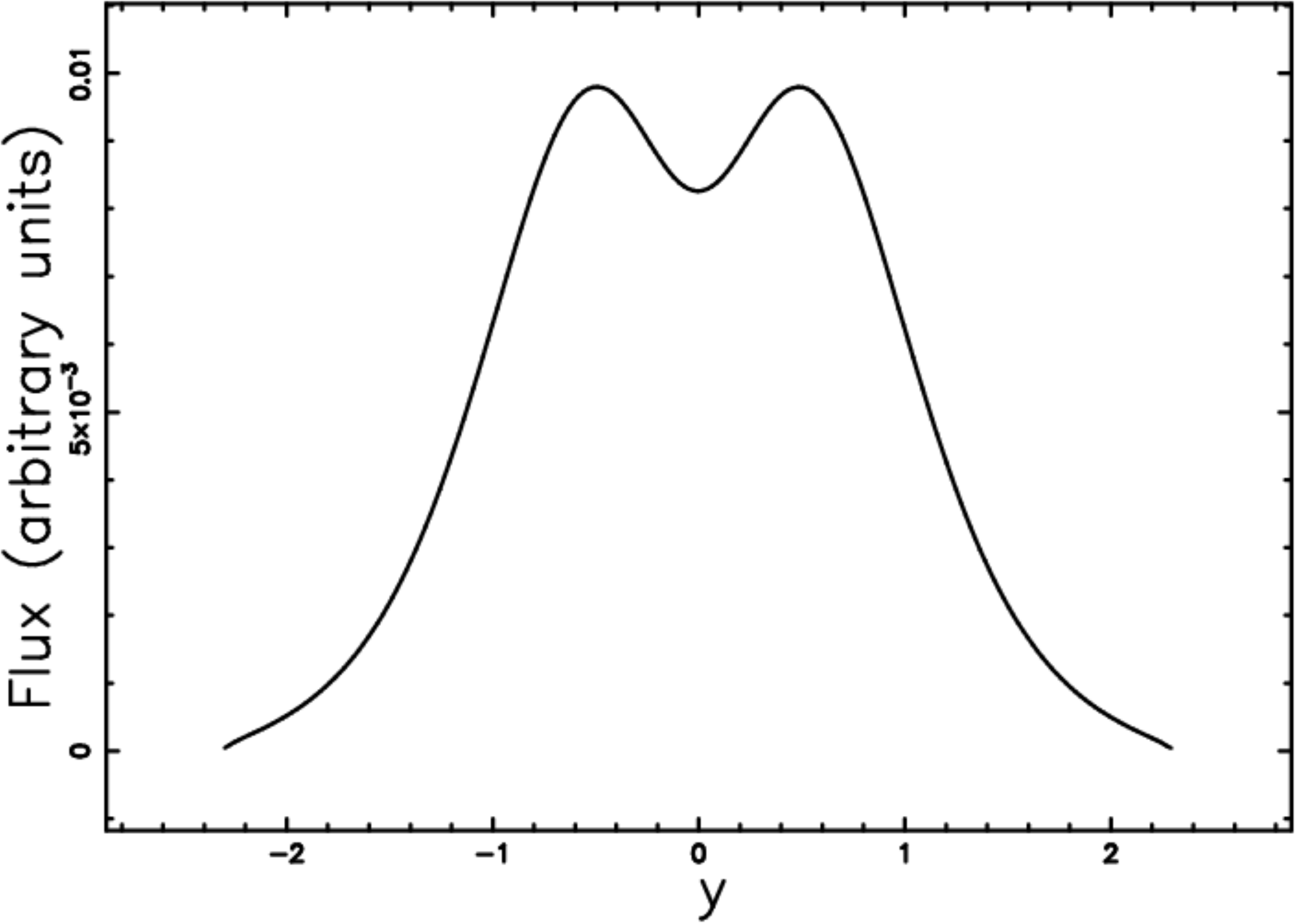}
  \end {center}
\caption {
 Intensity of radiation    ${\it I(y)}$ across the turbulent jet
 when  $z/d=44$ , 
       $k$ =0.54,
       $A~$ = 0.414 
       and  
$\alpha_{deg}=10.64$ .
          }%
    \label{integrale_turb}
    \end{figure}

This integral has an analytical solution but it is
complicated and 
therefore Figure~\ref{integrale_turb} 
only shows 
the numerical integration.
The numerical integration gives a characteristic shape 
on the top of the blob
which is  called   the "valley on the top" 
and  the observational counterpart 
of this new effect  will be discussed  in 
Section~\ref{blob}.
The maximum  of this integral is at the point  
$y \approx 0.49135{b_{1/2}}$ and the 
value of intensity at  the maximum is $1.18$ times
the value at the point $y=0$ ( the center of the jet).

\subsection{Profiles of the pipe fluids}

The intensity of emission in pipe fluids   
according to formula~(\ref{powern}) is 
\begin{equation}
I(y)     \sim 
\int _{0}^{\sqrt {{a}^{2}-{y}^{2}}}
\frac
{
{\it 2  \nu_n }\, \left [ {v_{\bot,max}}  \left( -{\frac {-a+\sqrt {{x}^{2}+{y}^{2}}}{a}}
 \right) ^{{n}^{-1}} \right] ^{2}
}
{
{n}^{2} \left( a-\sqrt {{x}^{2}+{y}^{2}} \right) ^{2}
}
{dx}
\quad  .
\label{iyn} 
\end{equation}

\begin{figure}
  \begin{center}
\includegraphics[width=10cm]{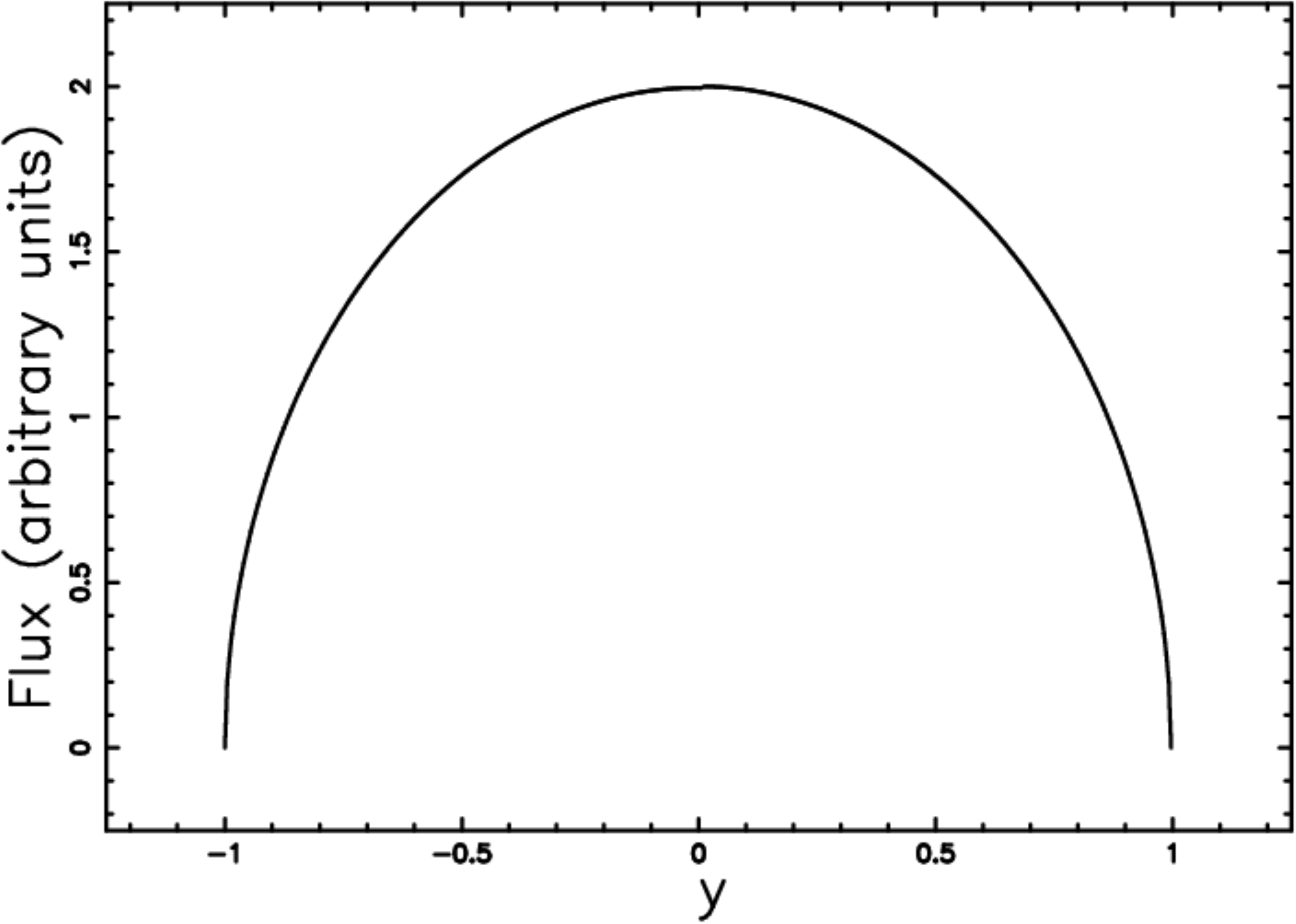}
  \end {center}
\caption {
 Intensity of radiation    ${\it I(y)}$ across the pipe fluid  jet,  
 when  $\nu_n=1$   , $a=1$ , $ {v_{\bot,max}}=1 $  and 
 $n=1$    .
          }%
    \label{integrale_n}
    \end{figure}

The previous  integral has an analytical solution in the case 
$n=1$ 
\begin{equation}
I(y) =   
2\,{\frac {{\it \nu_n}\,{{\it {v_{\bot,max}}   }}^{2}\sqrt {{a}^{2}-{y}^{2}}}{{a}^{
2}}}
\quad ,
\end{equation}
which  is reported in 
Figure~\ref{integrale_n}.

\begin{figure}
  \begin{center}
\includegraphics[width=10cm]{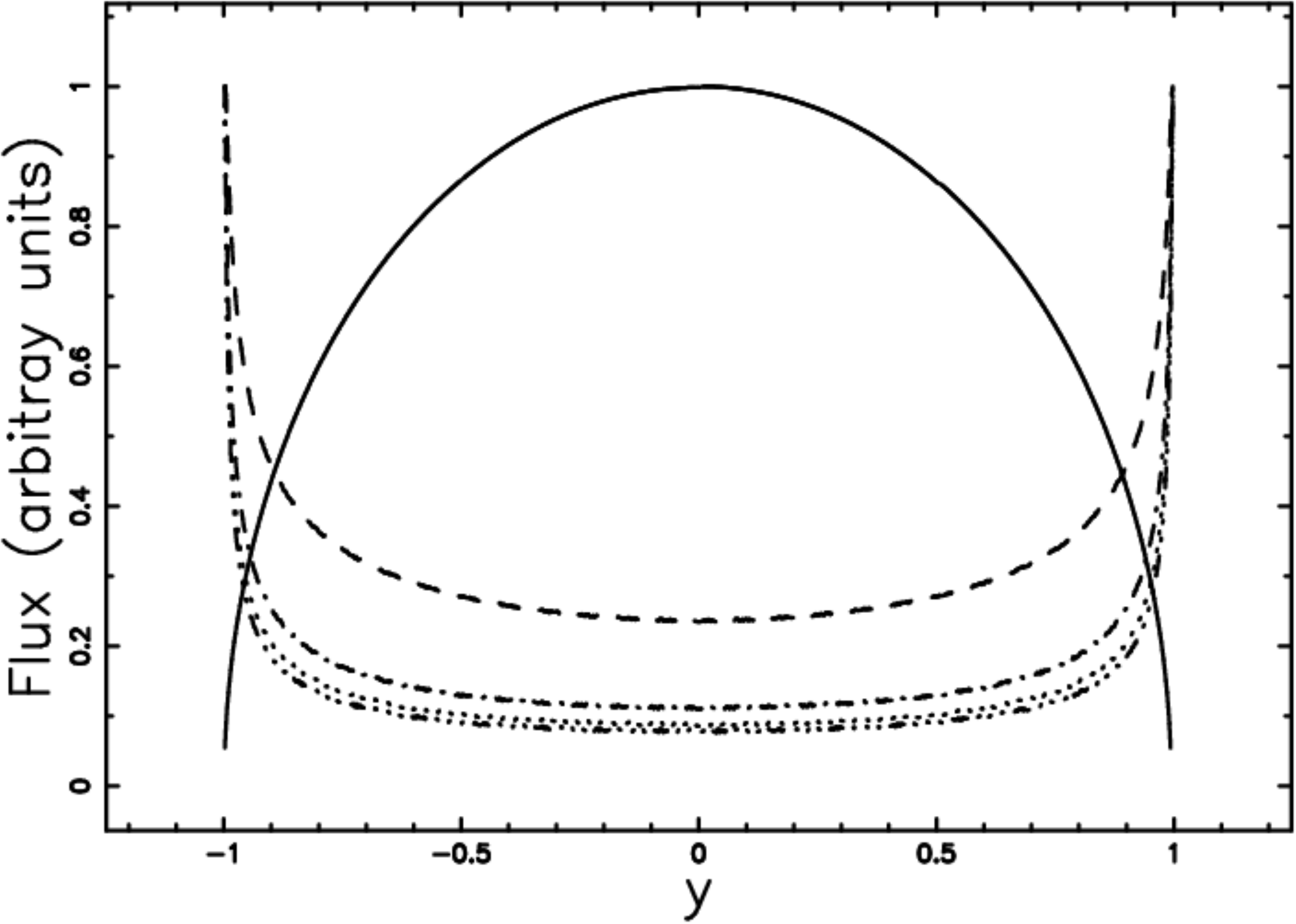}
  \end {center}
\caption {
 Intensity of radiation    ${\it I(y)}$ across the pipe fluid  jet
 normalized to  1    ,  
 when  $\nu_n=1$   , $a=1$ , $ {v_{\bot,max}}=1 $  
 and   
 $n=1$                     (full line        ),
 $n=2$                     (dashed           ),
 $n=3$                     (dot-dash-dot-dash),
 $n=4$                     (dotted)  and 
 $n=5$                     (dash-dot-dot-dot) .
          }%
    \label{integrale_n_new}
    \end{figure}

When  $n \neq 1 $
the integral cannot be solved in terms of any of the standard mathematical
functions

and  Figure~\ref{integrale_n_new} reports the  numerical integration
.

\subsection{Profiles of non-Newtonian fluid}

\begin{figure}
  \begin{center}
\includegraphics[width=10cm]{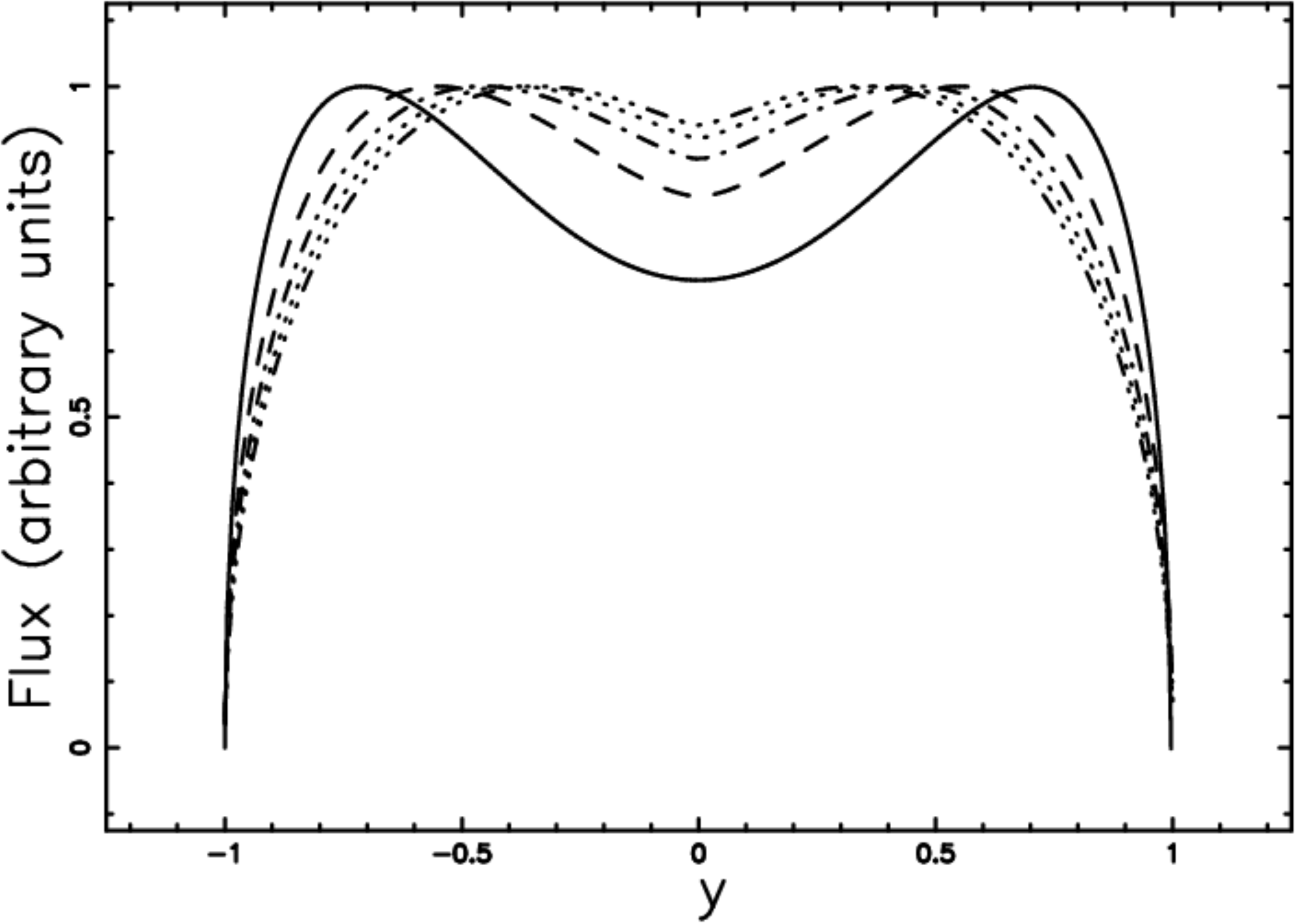}
  \end {center}
\caption {
 Intensity of radiation    ${\it I(y)}$ 
 across the {\it non-Newtonian} jet  
 normalized to  1
 when  $\nu_{non}$=1 , $a=1$ and $u_{max}$=1, 
 and   
 $m=1$                     (full line        ),
 $m=2$                     (dashed           ),
 $m=3$                     (dot-dash-dot-dash),
 $m=4$                     (dotted)  and 
 $m=5$                     (dash-dot-dot-dot) .
          }%
    \label{integrale_non}
    \end{figure}

The intensity of emission  
in a  {\it non-Newtonian fluid}
according to formula~(\ref{powernon}) is 
\begin{equation}
I(y)     = 
\int _{0}^{\sqrt {{a}^{2}-{y}^{2}}}
\frac
{
2\,{\it \nu_{non}}\,{{\it u_{max}}}^{2} \left[  \left( {\frac {\sqrt {{x}^{2}+
{y}^{2}}}{a}} \right) ^{{\frac {m+1}{m}}} \right] ^{2} \left( m+1
 \right) ^{2}
}
{
{m}^{2} \left( {x}^{2}+{y}^{2} \right)
}
{dx}
\quad .
\label{iynon} 
\end{equation}
Inserting $\nu_{non}$=1 , $a=1$ and $u_{max}$=1 for 
the 
sake of simplicity,
the previous  integral has the 
analytical solution  
\begin{eqnarray}
I(y) =  &                          ~\nonumber                    \\
2\, \left( {y}^{2} \right) ^{{\frac {m+1}{m}}}\sqrt {1-{y}^{2}}
{\2F1(1/2,1-{\frac {m+1}{m}};\,3/2;\,-{\frac {1-{y}^{2}}{{y}^{2}}})}
 \left( m+1 \right) {y}^{-2}{m}^{-1}+  \nonumber\\
2\, \left( {y}^{2} \right) ^{{
\frac {m+1}{m}}}\sqrt {1-{y}^{2}}
{\2F1(1/2,1-{\frac {m+1}{m}};\,3/2;\,-{\frac {1-{y}^{2}}{{y}^{2}}})}
 \left( m+1 \right) {m}^{-2}{y}^{-2}
   &~
\end{eqnarray}
where ${\2F1(a,b;\,c;\,z)}$ is a regularized hyper-geometric function,
see~\citet{Abramowitz1965},\citet{Seggern1992} and \citet{Thompson1997}.
Figure~\ref{integrale_non}
reports 
the results of the 
numerical integration.

\subsection{Comparison with observations}

\label{blob}
\begin{figure}
  \begin{center}
\includegraphics[width=10cm]{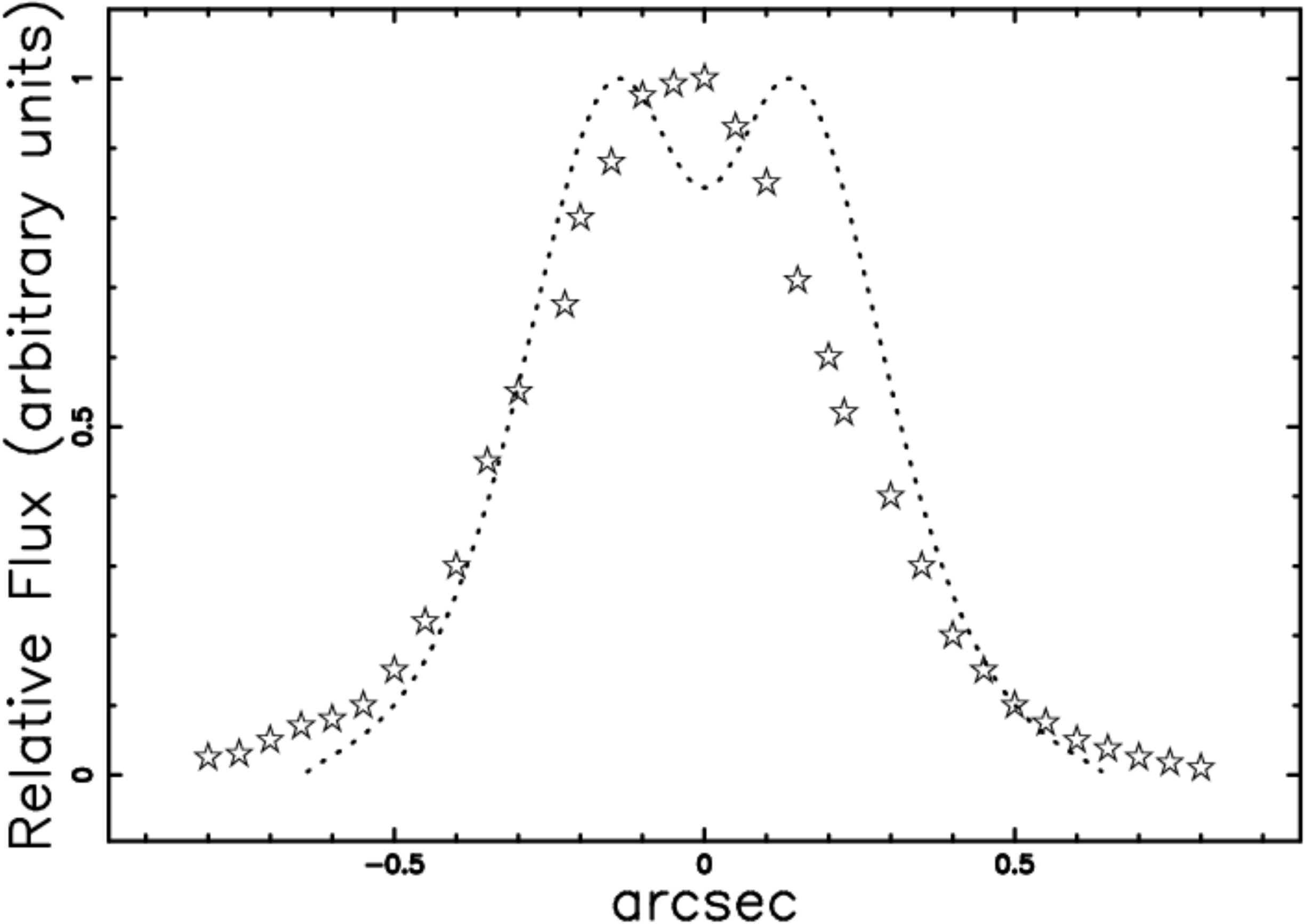}
  \end {center}
\caption {
 X-intensity of knot D-E in M87 (empty stars ) and 
 theoretical intensity of radiation for turbulent fluids
    ${\it I(y)}$  (dotted line)
  when  $z/d=44$ , 
        $k$ =0.54,
        $A~$ = 0.414 
        and  
        $\alpha_{deg}=10.64$ .
          }%
    \label{blob_turb}
    \end{figure}

\begin{figure}
  \begin{center}
\includegraphics[width=10cm]{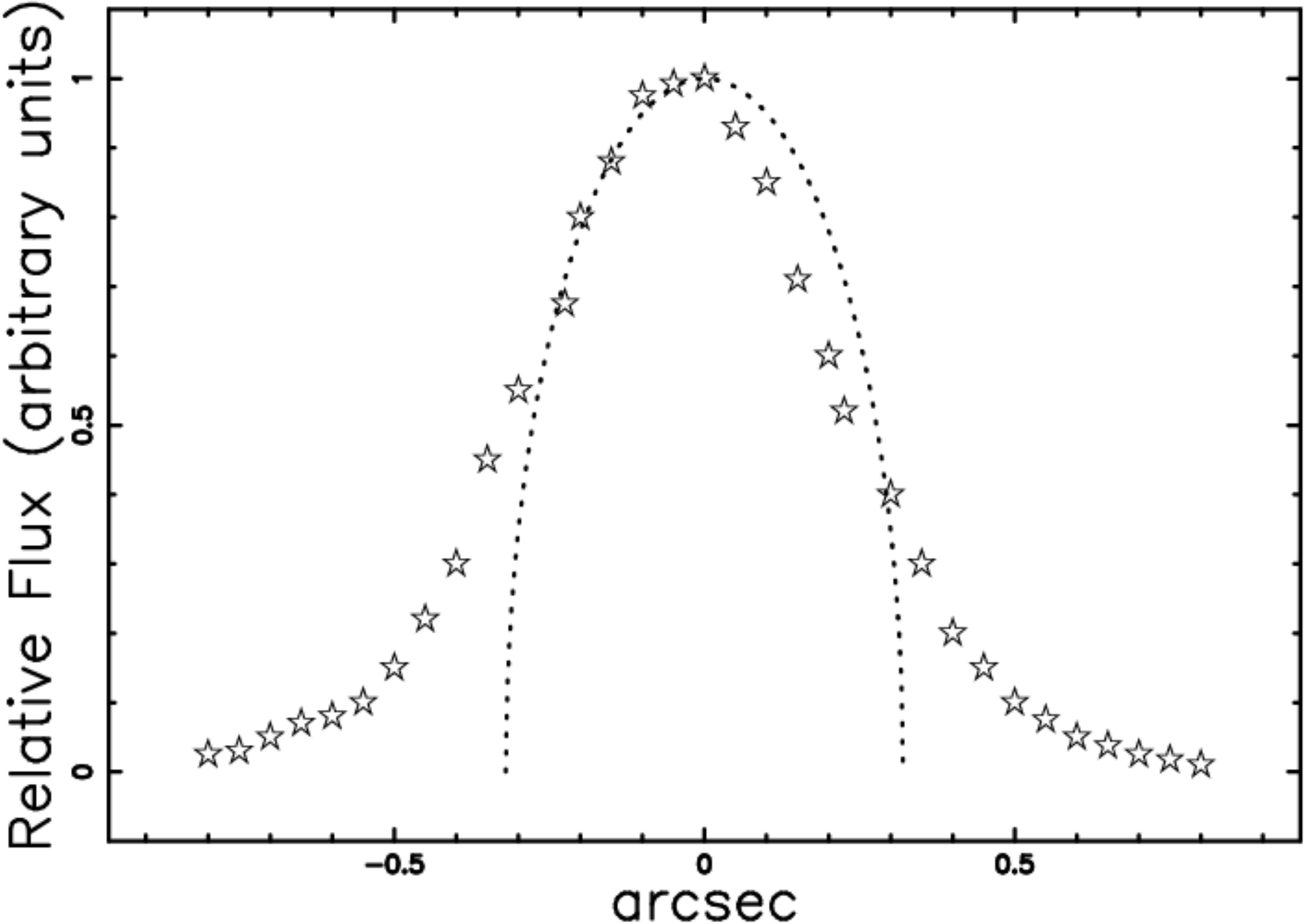}
  \end {center}
\caption {
  X-intensity of knot D-E in M87 (empty stars ) and 
  theoretical intensity of radiation  for a  pipe-fluid   ${\it I(y)}$
  (dotted line)
  when  $\nu_n=1$   , $a=1$ , $ {v_{\bot,max}}=1 $  and 
  $n=1$           .
          }%
    \label{blob_n}
    \end{figure}

\begin{figure}
  \begin{center}
\includegraphics[width=10cm]{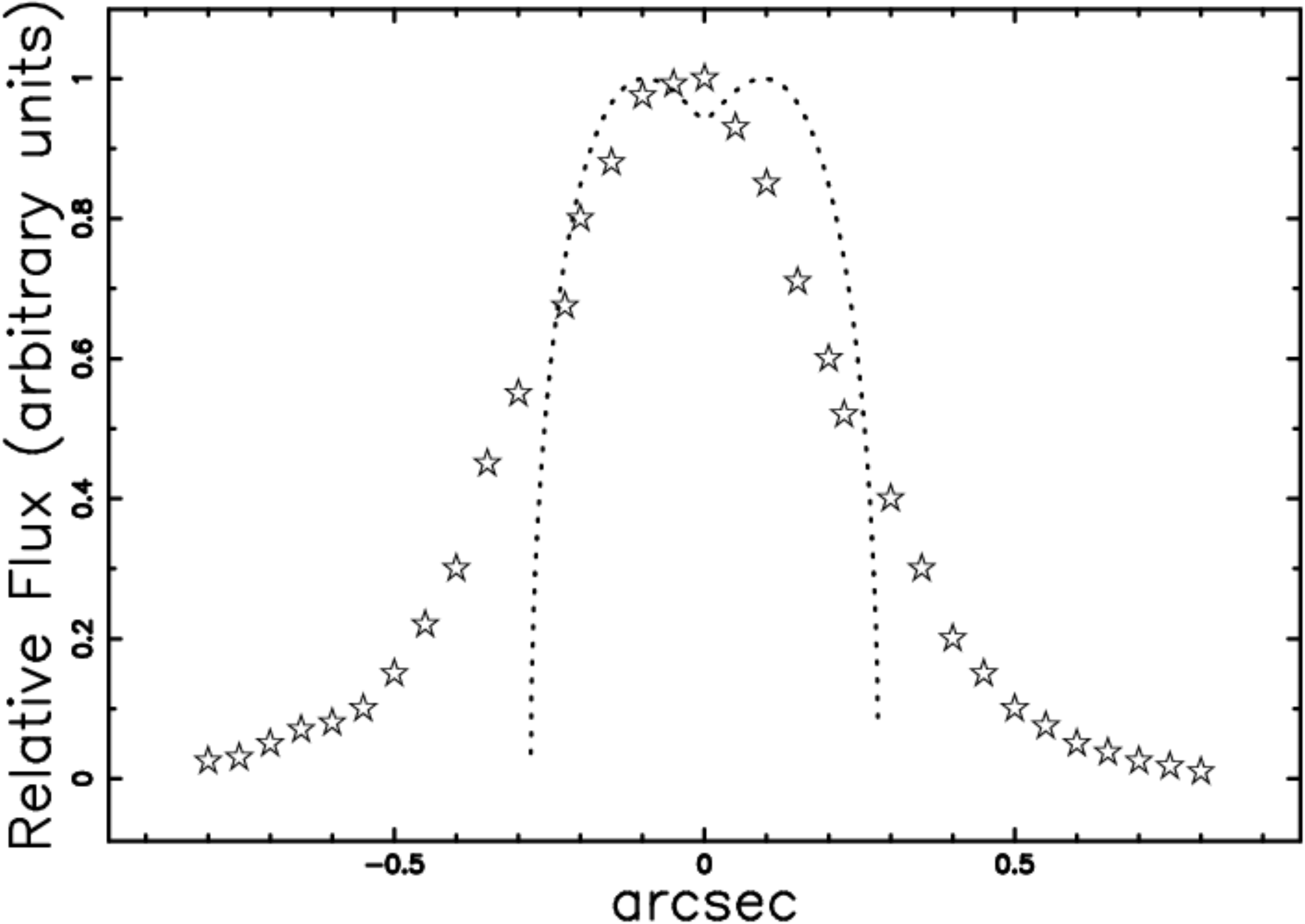}
  \end {center}
\caption {
 X-intensity of knot D-E in M87 (empty stars ) and 
 theoretical intensity of radiation  for a non 
 {\it non-Newtonian}  fluid   ${\it I(y)}$  (dotted line)
 when  $\nu_{non}$=1 , 
   $a=1$ , $u_{max}$=1, 
   and $m=5$ (dotted line).
          }%
    \label{blob_nonnewton}
    \end{figure}

This paragraph   starts by checking the profile in the direction 
perpendicular to the jet of M87 , in particular to the X-component
of knot D-E in Figure~2 by~\citet{Perlman2005}.
The data of the observations are reported in Figure~\ref{blob_turb}
together with the turbulent jet  profile scaled in order so 
that the value of half theoretical and observed intensity are
approximately equal.
Figures~\ref{blob_n} and~\ref{blob_nonnewton}  report the comparison 
of   pipe-fluid  and   
{\it non-Newtonian} fluid respectively.
 
Another  interesting situation is represented 
by the four blobs in 3C273 as imaged at the frequency of 5GHz 
by the VSOP (VLBI ( Very Long Baseline Interferometry) Space Observatory
Programme) .
Here  the focus is on blob A which   presents 
a symmetry around the center of the blob
and on the "valley on the top" .

\begin{figure}
  \begin{center}
\includegraphics[width=10cm]{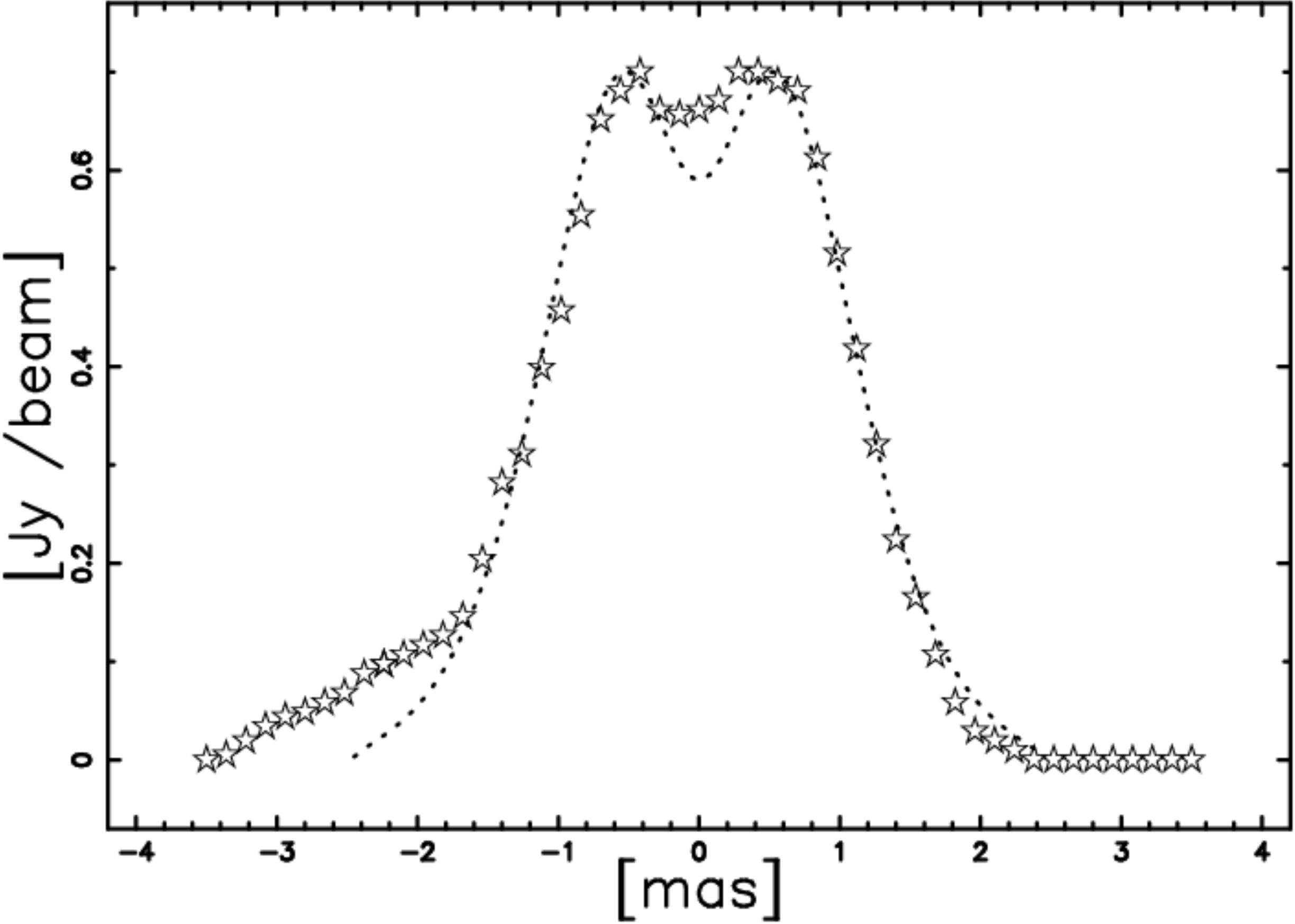}
  \end {center}
\caption {
 VSOP at the frequency of 5GHz of region A  in 3C273  (empty stars ) and 
 theoretical intensity of radiation for turbulent fluids
 , ${\it I(y)}$, (dotted line) 
  when  $z/d=44$ , 
        $k$ =0.54,
        $A~$ = 0.414 
        and  
        $\alpha_{deg}=10.64$ .
          }%
    \label{blob_turb_3C273}
    \end{figure}

The comparison  between
observed and theoretical-turbulent   "valleys  on the top" 
, see  Figure~\ref{blob_turb_3C273},
shows  that 
the theoretical depression is bigger than
the one observed.
In the case of {\it non-Newtonian}  fluid  
$m=5$  produces a perfect agreement between observed 
and theoretical depressions  for  blob A of 3C273,
see Figure~\ref{blob_nonnewton_3C273}.
\subsection{Sobel Filter}

\begin{figure}
  \begin{center}
\includegraphics[width=10cm]{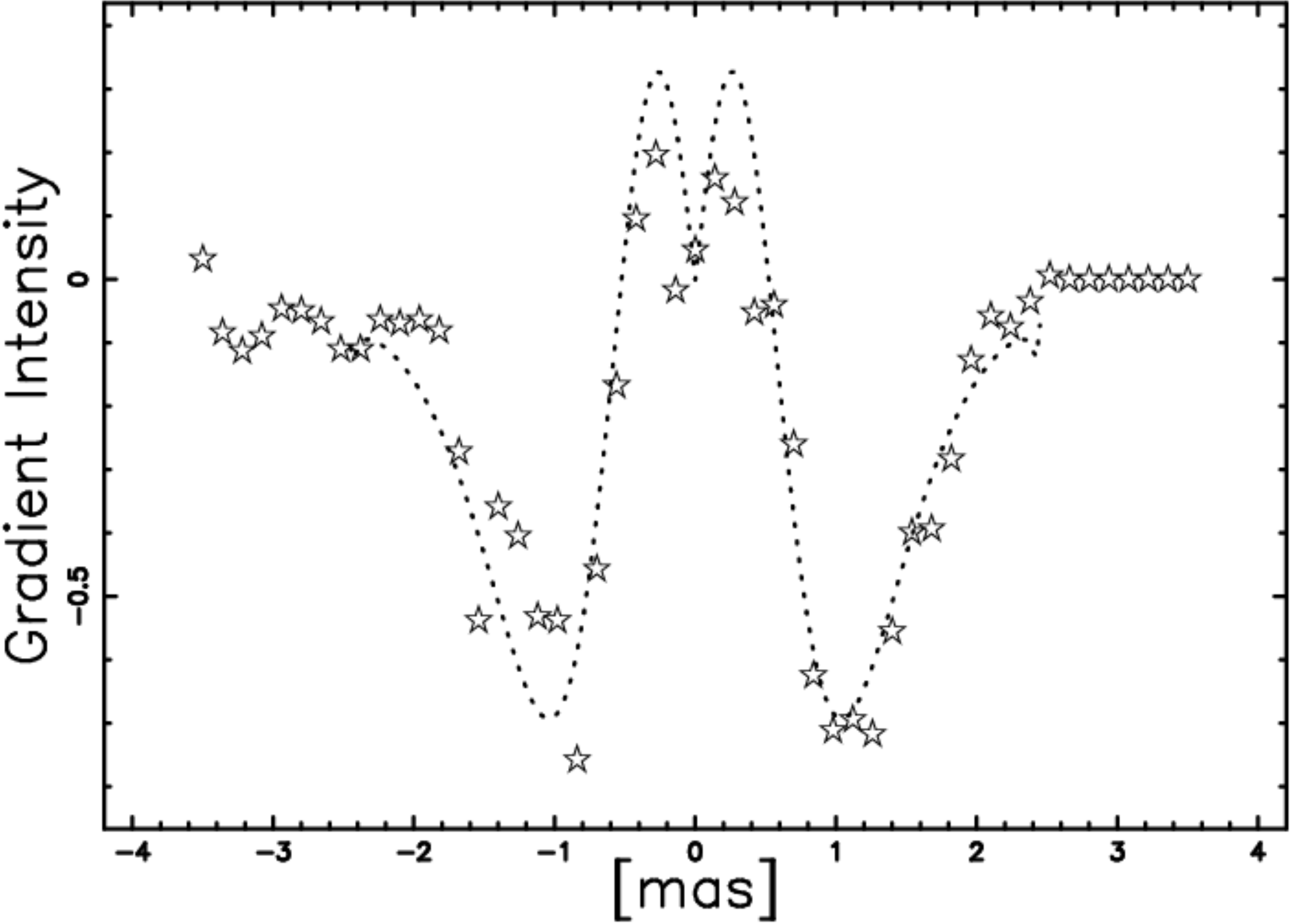}
  \end {center}
\caption {
Gradient of the intensity          (Sobel filter)  for 
VSOP-data  at  the frequency of  5GHz of region A  in 3C273  (empty stars )
and 
gradient of theoretical intensity (Sobel filter) 
of radiation for turbulent fluids  (dotted line ).
Parameters as  in Figure~\ref{blob_turb_3C273}.
          }%
    \label{blob_turb_3C273_sobel}
    \end{figure}

\begin{figure}
  \begin{center}
\includegraphics[width=10cm]{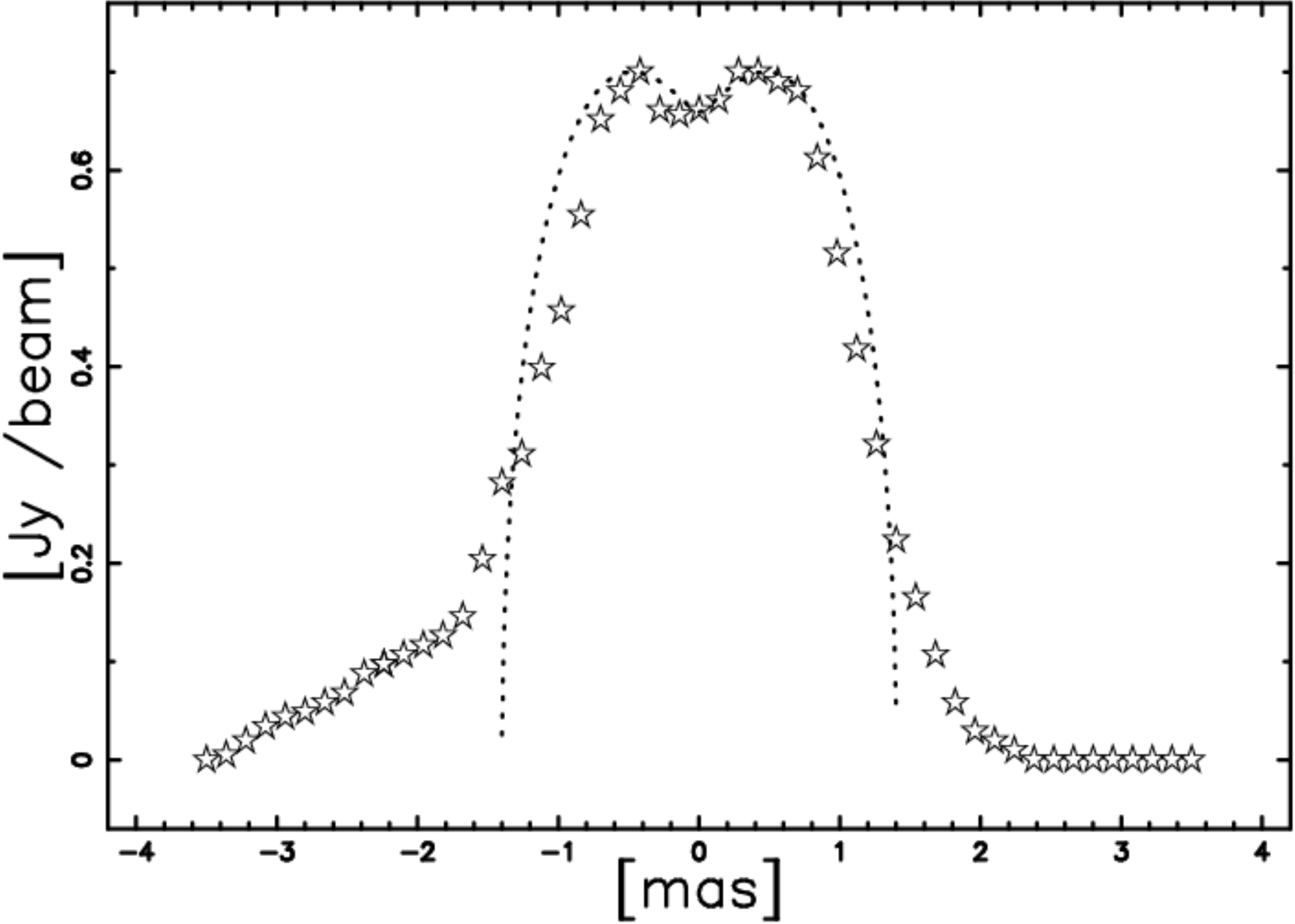}
  \end {center}
\caption {
  VSOP at  the frequency of   5GHz    of region A  in 3C273 (empty stars ) and 
 theoretical intensity of radiation  for a non 
 {\it non-Newtonian  fluid}   ${\it I(y)}$  (dotted line)
   when  $\nu_{non}$=1 , 
   $a=1$               ,
   $u_{max}$=1         , 
   and $m=5$ (dotted line).
          }%
    \label{blob_nonnewton_3C273}
    \end{figure}

The Sobel filter , see~\citet{Pratt_1991} , computes the
spatial gradient of the intensity of radio-galaxies ,
see , for example,  
Figure~4b and  Figure~26b in~\citet{Laing_2006},
which  represent  respectively 3C296 and 3C31.
The Sobel filter is now applied in the following way .
${\bf S}$ is the   the source image , 
${\bf G_x}$ and ${\bf G_y}$ 
are 
the two images which at each point contain the horizontal 
and vertical  first derivative of ${\bf S} $.
At each point in the image the resulting gradient approximations
can be combined to give the gradient magnitude 
${\bf G}$ 
\begin{equation}
{ \bf G} = \sqrt {{\bf G_x} ^2  + {\bf G_y}^2} 
\quad ,
\label{gradient}
\end{equation}
where 
\begin{equation}
{\bf G_x} =
\left[ \begin {array}{ccc}
1&0&-1\\\noalign{\medskip}2&0&-2\\\noalign{\medskip}1&0&-1\end {array}
\right] 
\ast {\bf S}
\end {equation}
and 
\begin{equation}
{\bf G_y} =
 \left[ \begin {array}{ccc} 1&2&1\\\noalign{\medskip}0&0&0
\\\noalign{\medskip}-1&-2&-1\end {array} \right] 
\ast {\bf S}
\end {equation}
where  $\ast$ denotes the 2-dimensional convolution operation.
The gradient  direction, $\Theta$ , is
\begin{equation}
\Theta = \arctan \left ( \frac  {{\bf G_y}}  {{\bf G_x}} \right )
\quad .
\end{equation}
Here we have implemented the Sobel filter to 1D  cuts 
in intensity of 3C273; 
Figure~\ref{blob_turb_3C273_sobel} reports the theoretical 
and the observed Sobel filter.

\section{2D maps}

An algorithm  is now built  , which is  able to simulate 
the synchrotron emission of the jet
given the basic assumption that the emissivity $\epsilon$
scales as the power released in turbulence for a turbulent fluid.
The integral operation that gives the intensity on a plane
is   performed through a simple addition as given 
by formula~(\ref{transfer_sum}).
The simplest case , a jet perpendicular to the line of sight  
is now analyzed.
\label{2D}

\begin{figure}
  \begin{center}
\includegraphics[width=10cm]{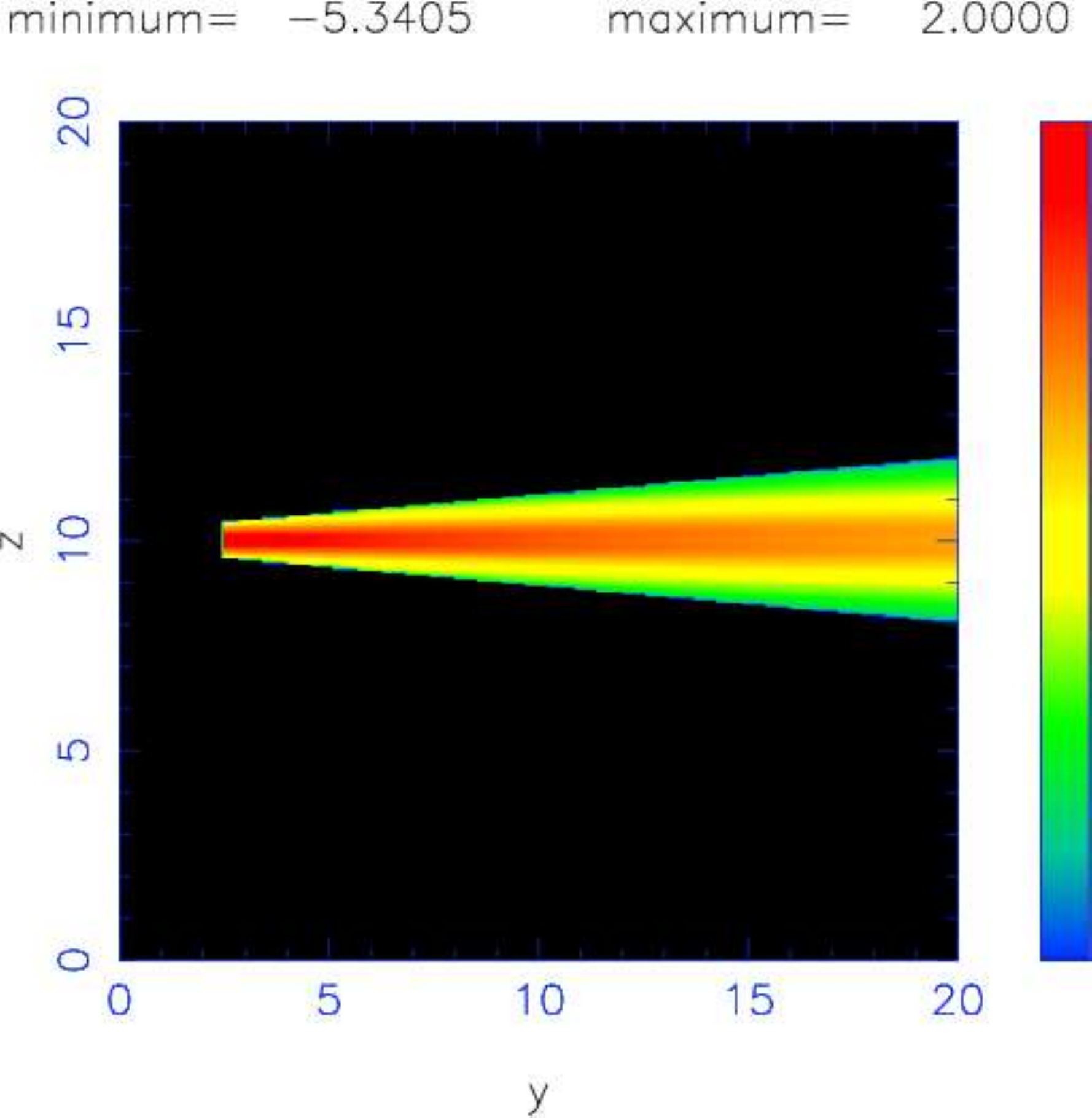}
  \end {center}
\caption {
Theoretical 2D map of the 
decimal logarithm of  the
intensity of emission
over the  astrophysical length of 20$kpc$
corresponding to the radio-galaxy 
NGC4061  at 21  cm ,
see  Figure~5
in \citet{Zaninetti2007_b}~.
The length  expressed in the nozzle's diameter units is $z/d=44$ , 
$k$ =0.54,
$A~$ = 0.414 
and  
$\alpha_{deg}=10.64$ .
The integral operation is performed on a  cubic  grid 
of $400^3$ pixels.
          }%
    \label{NGC4061_log}
    \end{figure}

The symmetry here used is that 
the jet's  
main direction is 
perpendicular to the observer,
see Figure~\ref{NGC4061_log}.

\begin{figure}
  \begin{center}
\includegraphics[width=10cm]{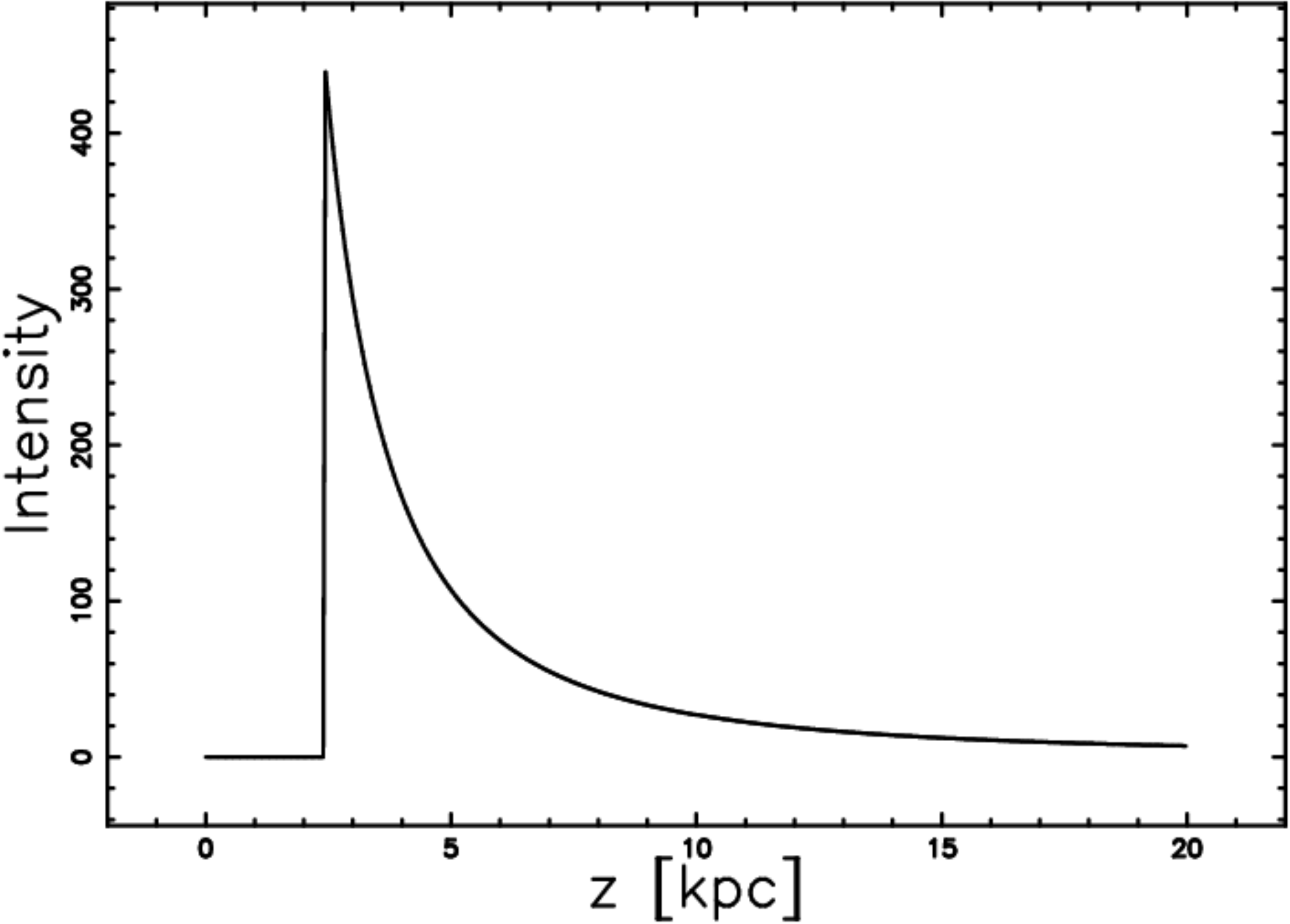}
  \end {center}
\caption {
Cut of  the intensity of emission along  the main direction 
at the center of the jet
as extracted from Figure~\ref{NGC4061_log} .
          }%
    \label{NGC4061_along}
    \end{figure}

\begin{figure}
  \begin{center}
\includegraphics[width=10cm]{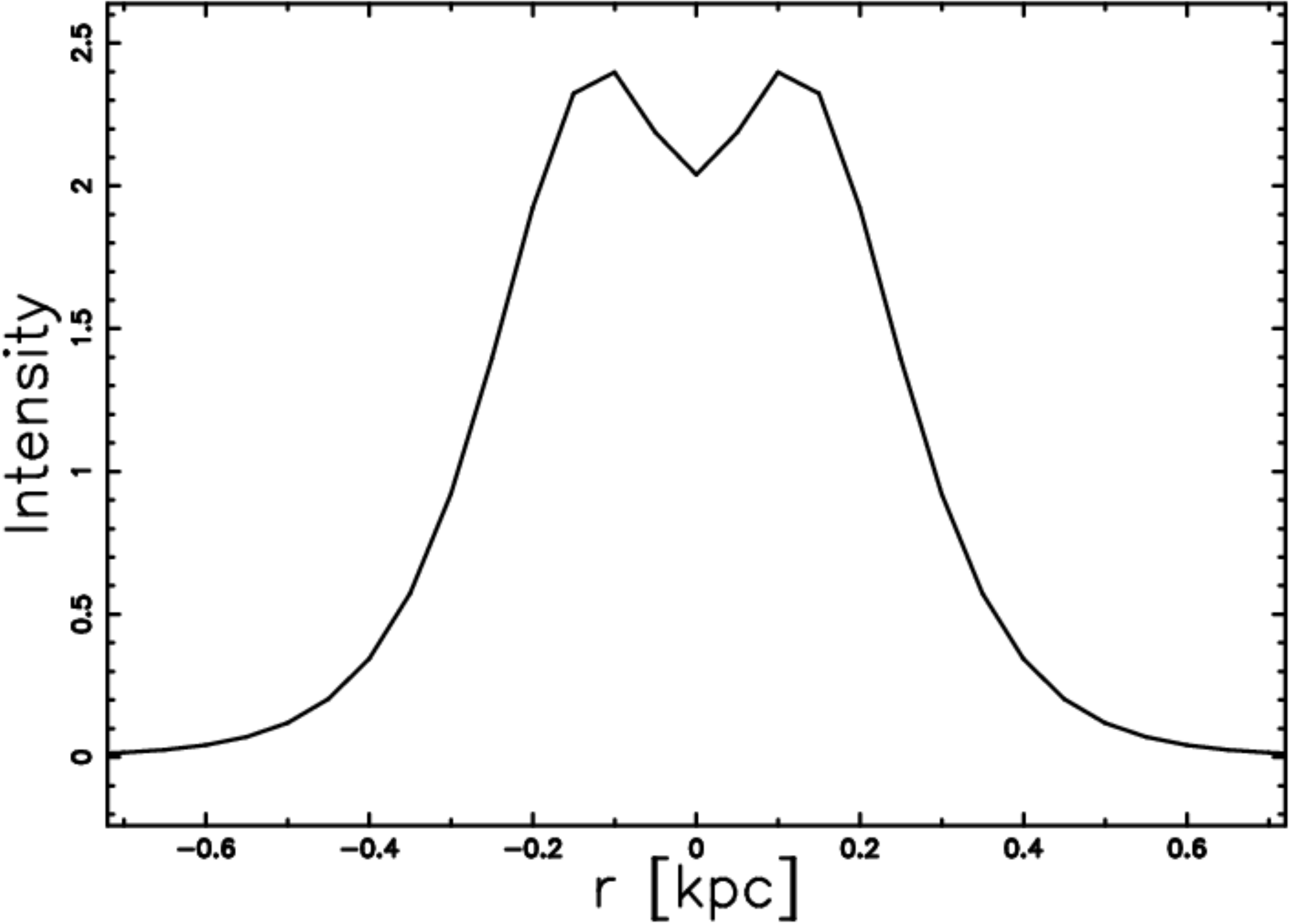}
  \end {center}
\caption {
Cut of   the intensity of emission in the transversal  direction 
, or $90^{\circ} $
in the middle of the jet 
as extracted from Figure~\ref{NGC4061_log} .
          }%
    \label{NGC4061_transverse}
    \end{figure}

In order to clarify  the interpretation of this theoretical 
2D color map 
Figure~\ref{NGC4061_along}  
 reports  the intensity profile along the main direction  
and the transversal profile as computed in the middle 
of the jet is shown in Figure~\ref{NGC4061_transverse}. 

\begin{figure}
  \begin{center}
\includegraphics[width=10cm]{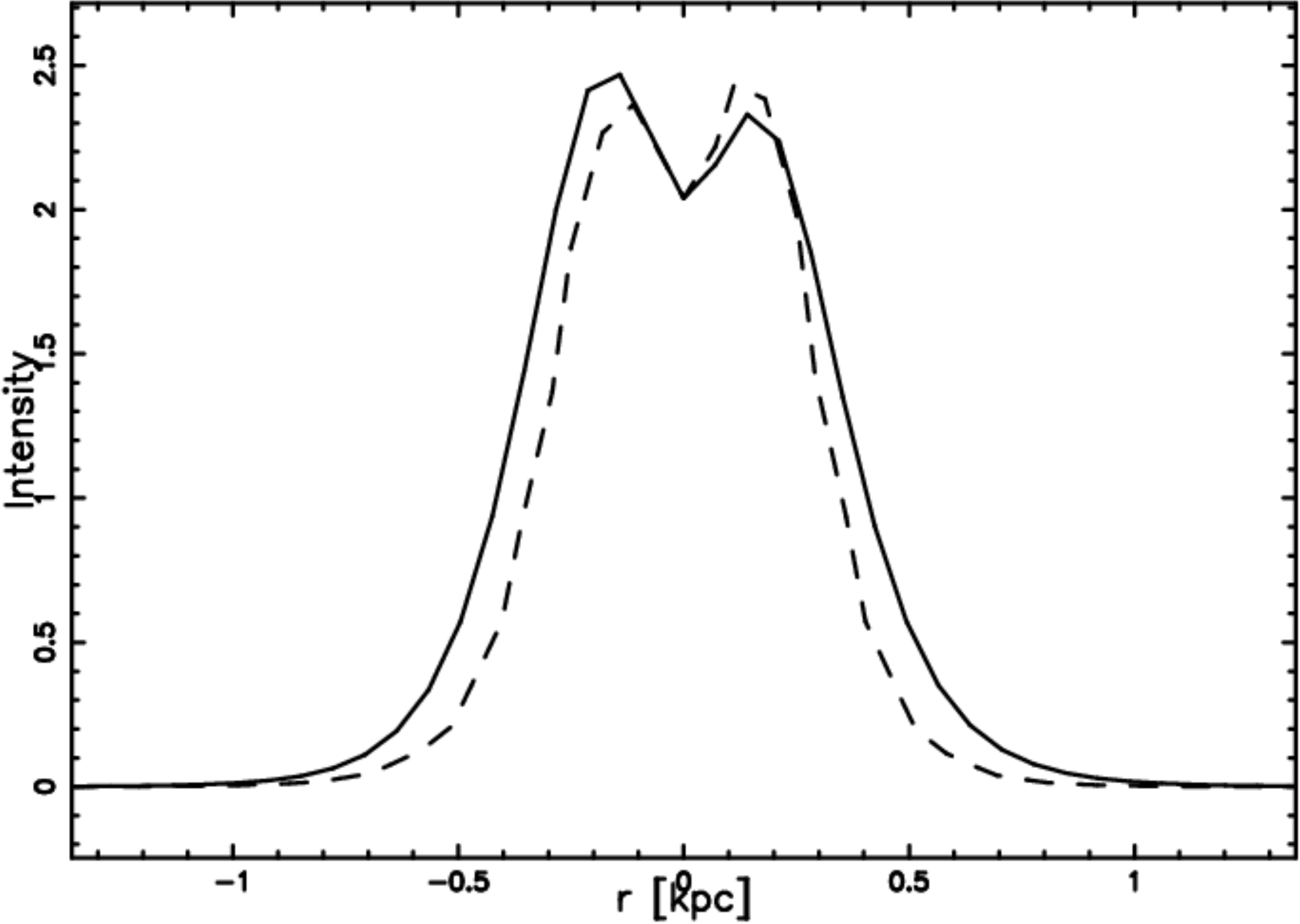}
  \end {center}
\caption {
Cut  of the intensity of emission   in the middle of the jet 
at      $45^{\circ}  $ (full   line) 
and  at $120^{\circ} $ (dashed line) 
as extracted from Figure~\ref{NGC4061_log} .
          }%
    \label{NGC4061_2_cuts}
    \end{figure}

The previous case represents a cut along a direction 
that forms an angle of $90^{\circ}  $   with  respect
to the jet's   main direction.
When the angles are different, the 
" valley on the top"   becomes asymmetric 
and the thickness of the cut increases , see 
Figure~\ref{NGC4061_2_cuts} where two profiles are reported.

\begin{figure}
  \begin{center}
\includegraphics[width=10cm]{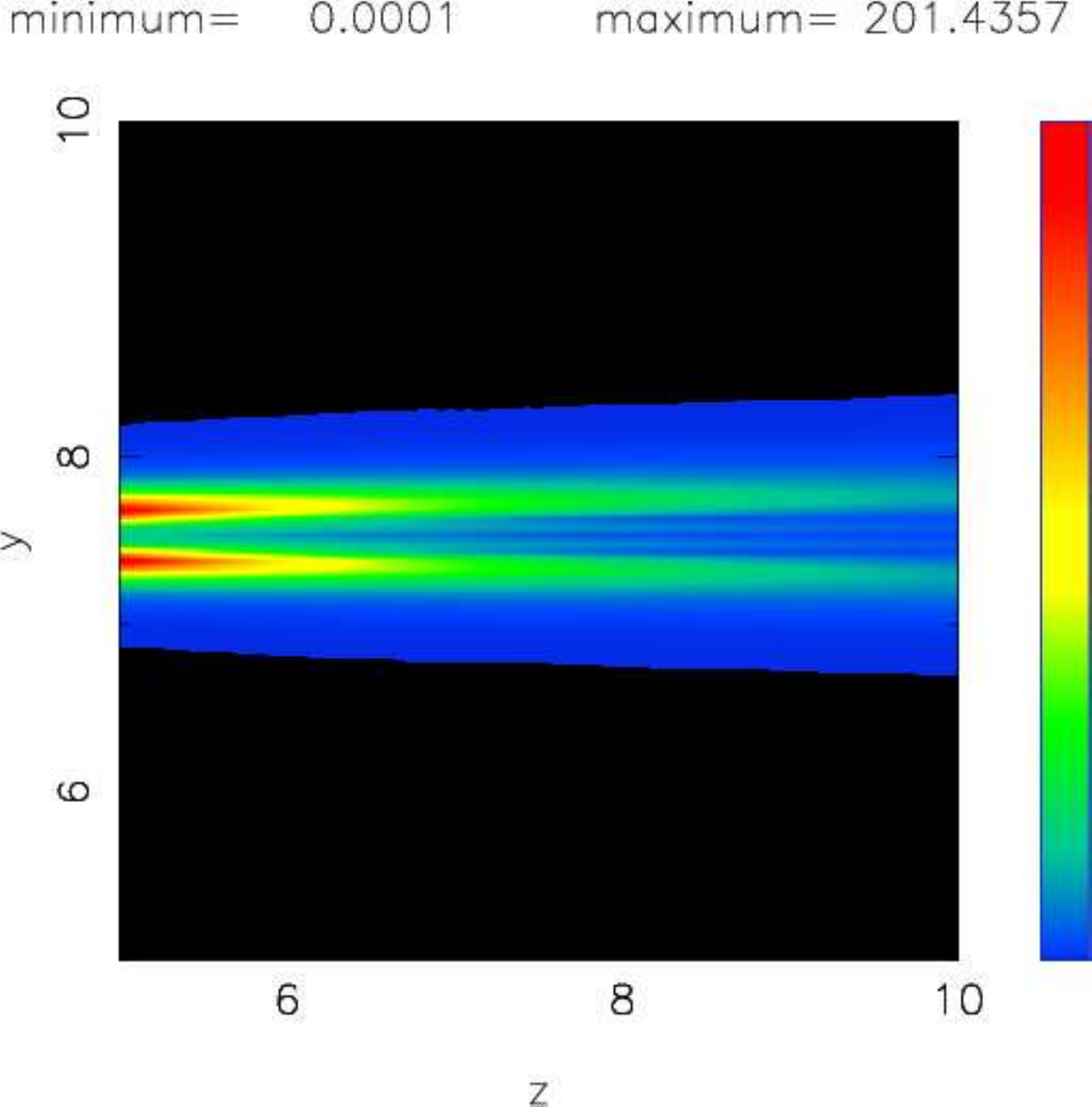}
  \end {center}
\caption 
{
Sobel filter of the intensity  of the   central part of  NGC4061.
The gradient magnitude is computed according to formula~(\ref{gradient}),
other parameters as in Figure~\ref{NGC4061_log}.
}
    \label{NGC4061_sobel_4}
    \end{figure}

The gradient magnitude 
of the previous image  is now plotted , 
see Figure~\ref{NGC4061_sobel_4}.

\section{Astrophysical applications}
\label{astro}

The following  reports the velocity , the equation of  motion 
, the power released in turbulence 
and  the flow rate of mass
when the 
astrophysical units are adopted.
The algorithm that allows us to build 
complex trajectories such as NGC6041 and 3C31 is then 
introduced.
The observed spectral index typical of the radio-galaxies 
is simulated.

\subsection{Astrophysical equations}

Here shown is  the astrophysical version of the previous derived  
equations for the turbulent jet.
Equation~(\ref{vlab}) allows us  to deduce the centerline velocity 
\begin{equation}
u_0(z) = 
 162027\,{\frac {\beta_0\,{\it d_1}}{\tan \left(  0.00872\,{
\it \alpha_{deg} } \right) z_1}}
\frac{Km}{sec}
\quad , 
\label{u_astro}
\end{equation}
where $\alpha_{deg}$ is the opening angle expressed in degrees,
$\beta_0$ is the initial velocity divided by the light velocity,
$d_1$ is the diameter of the nozzle in $kpc$ units 
and  
$z_1$ is the length of the jet in $kpc$ units. 
From the previous equation it is  possible to deduce 
the equation of  motion for a turbulent jet ,
\begin{equation}
z(t) =
 57.56\,\sqrt {{\frac {\beta_0\,{\it d_1}\,{\it t_7}}{\tan \left( 
 0.00872\,{\it \alpha_{deg}} \right) }}} kpc
\quad ,
\label{z_astro}
\end{equation}
where $t_7$ = $t/(10^7)$ \mbox {year}.
The radius of the turbulent jet is 
\begin{equation}
r(z) = [\frac {d_1}{2} + z(t) \tan (\frac {\alpha} {2} ) ] kpc
\quad ,
\label{r_astro}
\end{equation}
where $\alpha$ is the opening angle expressed in radians.
The power  released in the turbulent  cascade is
\begin{equation}
\epsilon(r,z)  = 
\frac {
2\, \left( \sqrt {2}-1 \right) ^{2}{{\it k}}^{2}{{\it b_{1/2}}}^{12}{A~
}^{2}{r}^{2}
}
{
\tan \left( 1/2\,\alpha \right)  \left( \sqrt {2}-1 \right) ^{3} \left( {{\it b_{1/2}}}^{2}+A~{r}^{2} \right) ^{6}{{\it b_{1/2}}}^{4}
}
(\frac {d_1}{z_1})^2 
\label{e_astro}
\quad ,
\end{equation}
where $z_1$ is the position of the jet expressed in $kpc$ units
, $A$ and $k$ are  defined in Section~\ref{turbulent} .

The flow rate of mass , see equation~(\ref{flow_practical}) ,
as expressed in these  astrophysical units 
is 
\begin{equation}
\dot {m}(z) =
8.0\,10^{11}{\it n_0}\,{\it z_1}\, \left[ \tan \left(
\frac {\alpha} {2}  \right)  \right] ^{3}\beta_0\,{\it d_1}\,
\frac { {\mathcal {M}}_{\sun}} {10^7  \mbox {year}} 
\quad  ,
\end{equation}
where  
$\rho_0=n_0 m$ , 
$m=1.4~m_{\mathrm {H}}$ ,
$n_0$  is the density expressed  in particles~$\mathrm{cm}^{-3}$ ,
$m_{\mathrm {H}}$ is the mass of the Hydrogen and
$ {\mathcal {M}}_{\sun}$ is the mass of the Sun.

\subsection{Astrophysical images}

\begin{figure}
  \begin{center}
\includegraphics[width=10cm]{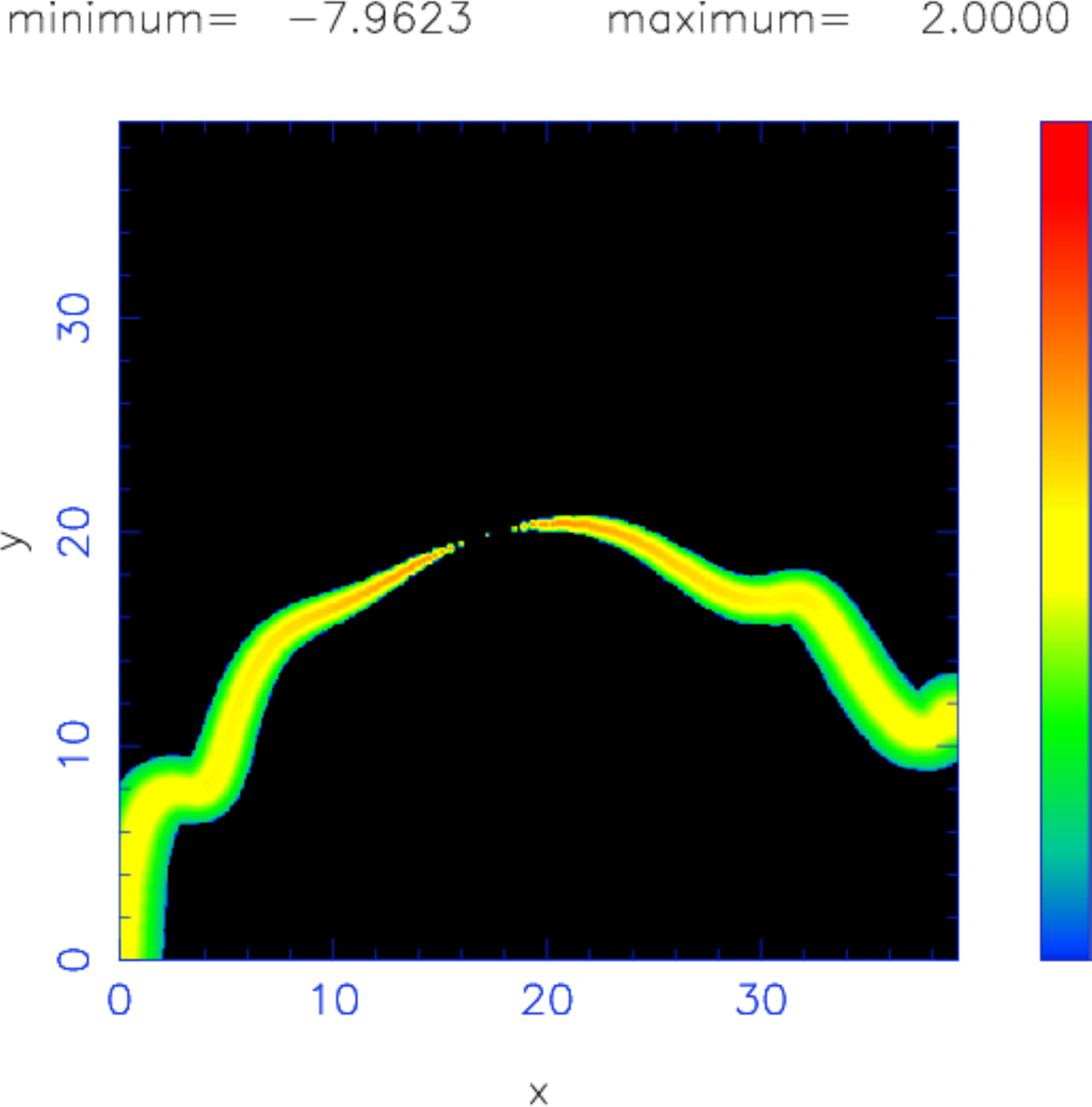}
  \end {center}
\caption {
Theoretical 2D map of the decimal logarithm of the
intensity of emission
representing the continuous three-dimensional trajectory of NGC4061:
 the three Eulerian angles 
 characterizing the point of view are $ \Phi $=80   $^{\circ }$
 , $ \Theta $=-85   $^{\circ }$
 and  $ \Psi $=-10   $^{\circ }$.
 The precession is characterized by the angle
  $ \Psi_{prec} $=  5   $^{\circ }$
 and by the angular velocity
 $ \Omega_{prec} $= 60.00 [$^{\circ}/10^7 \mathrm{year}$].
 The three Eulerian angles  which characterize   the jet are 
      $ \Phi_{j} $=  0   $^{\circ }$
 ,    $ \Psi_{j} $=  0   $^{\circ }$
 and  $ \Theta_{j} $= 90   $^{\circ }$.
 The angle of rotation of the galaxy is
 $ \alpha_{G} $=  0   $^{\circ }$.
 The physical parameters characterizing the jet motion
 are : $t_7$= 12.00~$10^7 \mathrm{year}$,
       $d_1$=  0.01~${\mathrm{kpc}}$ ,
       $\beta_0$ = 0.1               ,
       $\alpha_0$= 10.64   $^{\circ }$.
The length of the jet is  20.6 ${\mathrm{kpc}}$  and 
the velocity of the galaxy $v_z$= 117 ${\mathrm{km/s}}$ .
The integral operation is performed on cubic  grid 
of $500^3$ pixels.
          }%
    \label{NGC4061_complex_log}
    \end{figure}

The points belonging  to a trajectory , in the
following TP ,
are now inserted on a cubic lattice made by  $pixels^3$ points .
The intensity  in each point of a 2D grid
is computed according to the following rules.
\begin{enumerate}
\item In each point  of a 3D grid we compute the nearest  TP.
\item The concentration on a 3D grid is then computed because we know the
      distance between the grid point and the nearest TP. 
\item The intensity of radiation is then computed according 
      to  the integral operation as represented by 
      formula~(\ref{transfer_sum}). 
\end{enumerate}
A  first test can be done  on a radio galaxy already examined
in \citet{Zaninetti2007_b}, NGC4061.
The equation of motion is formula~(\ref{z_astro}),
 derived in the 
previous paragraph.
The other parameters necessary 
to produce bending and wiggling of the jet
as well as the Eulerian angles which 
represent the point of view
are  
reported in the caption of  Figure~\ref{NGC4061_complex_log}.

\begin{figure}
  \begin{center}
\includegraphics[width=10cm]{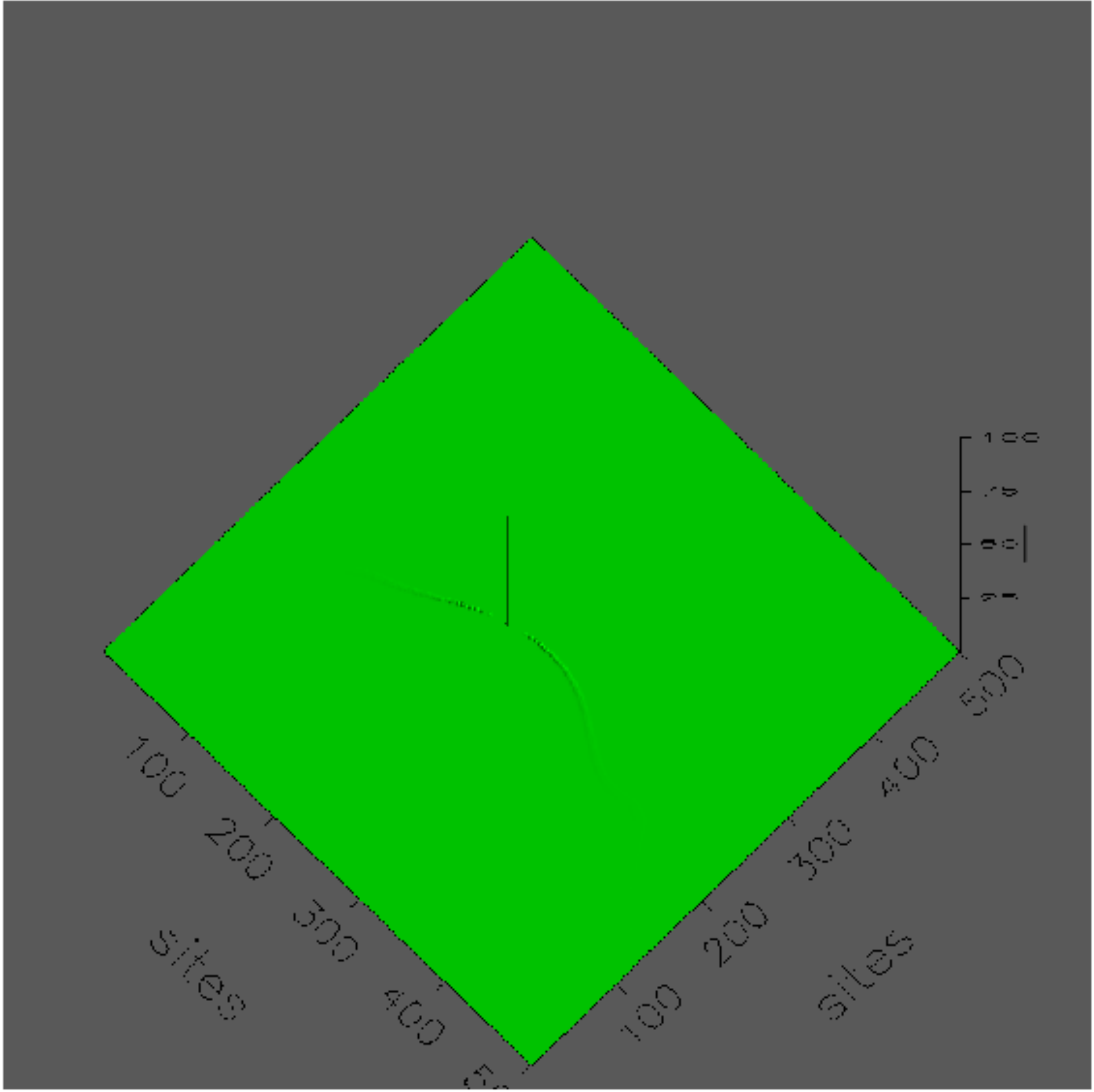}
  \end {center}
\caption {
Intensity  of NGC4061  represented
through a surface , parameters as  
in Figure~\ref{NGC4061_complex_log}. The  observer sees  the surface
at   an angle of 70$^{\circ }$.  
          }%
    \label{NGC4061_euler_surf}
    \end{figure}

\begin{figure}
  \begin{center}
\includegraphics[width=10cm]{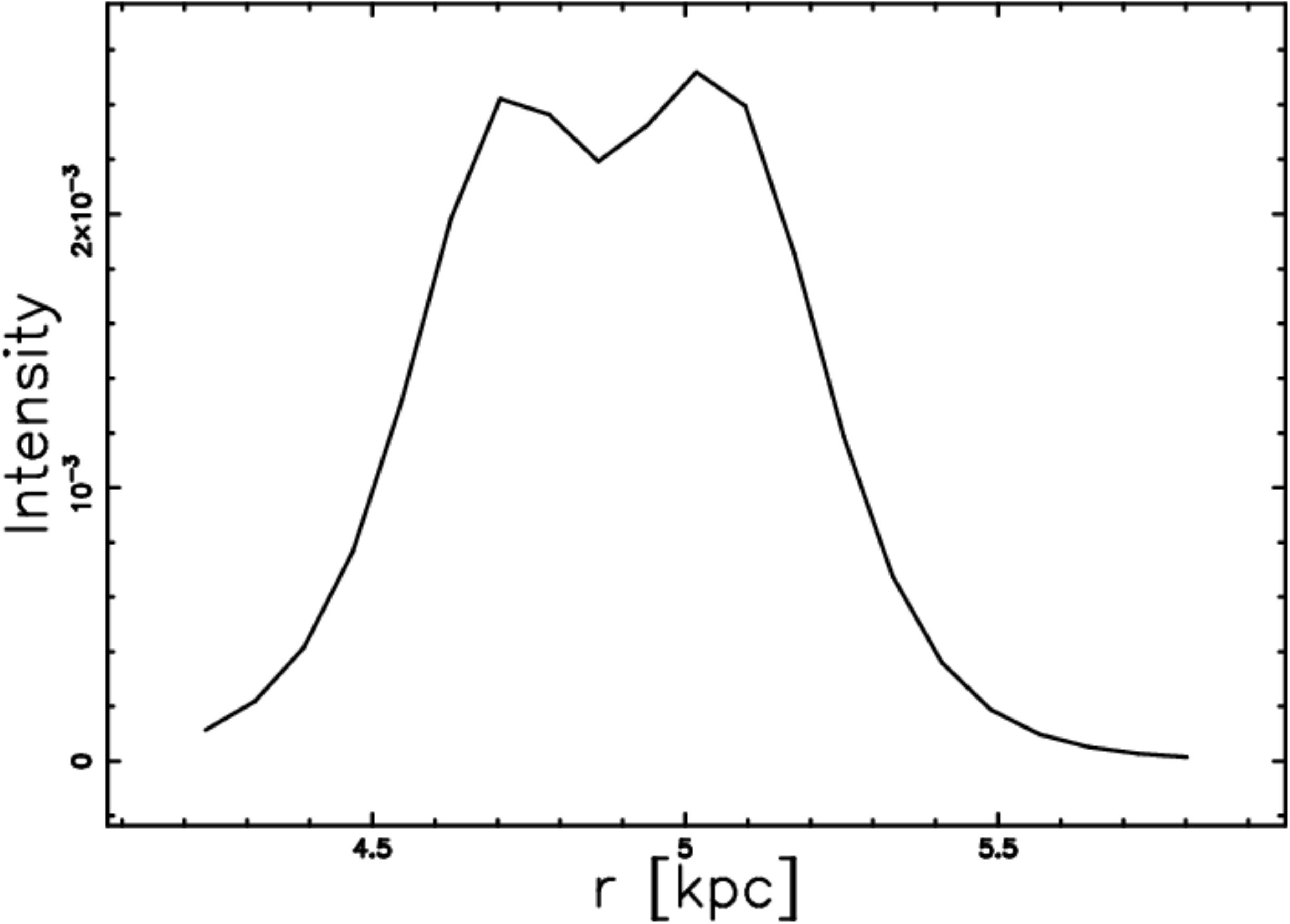}
  \end {center}
\caption {
Intensity  of NGC4061  represented
through a cut in the x-direction , parameters as  
in Figure~\ref{NGC4061_complex_log}. 
          }%
    \label{NGC4061_cut}
    \end{figure}

The phenomena of the "valley on the top" 
can be visualized through a surface rendering of the 2D intensity
 , see  Figure~\ref{NGC4061_euler_surf} ,
or through a 1D cut of the matrix that represents   
the intensity , see  Figure~\ref{NGC4061_cut}~.

\begin{figure}
  \begin{center}
\includegraphics[width=10cm]{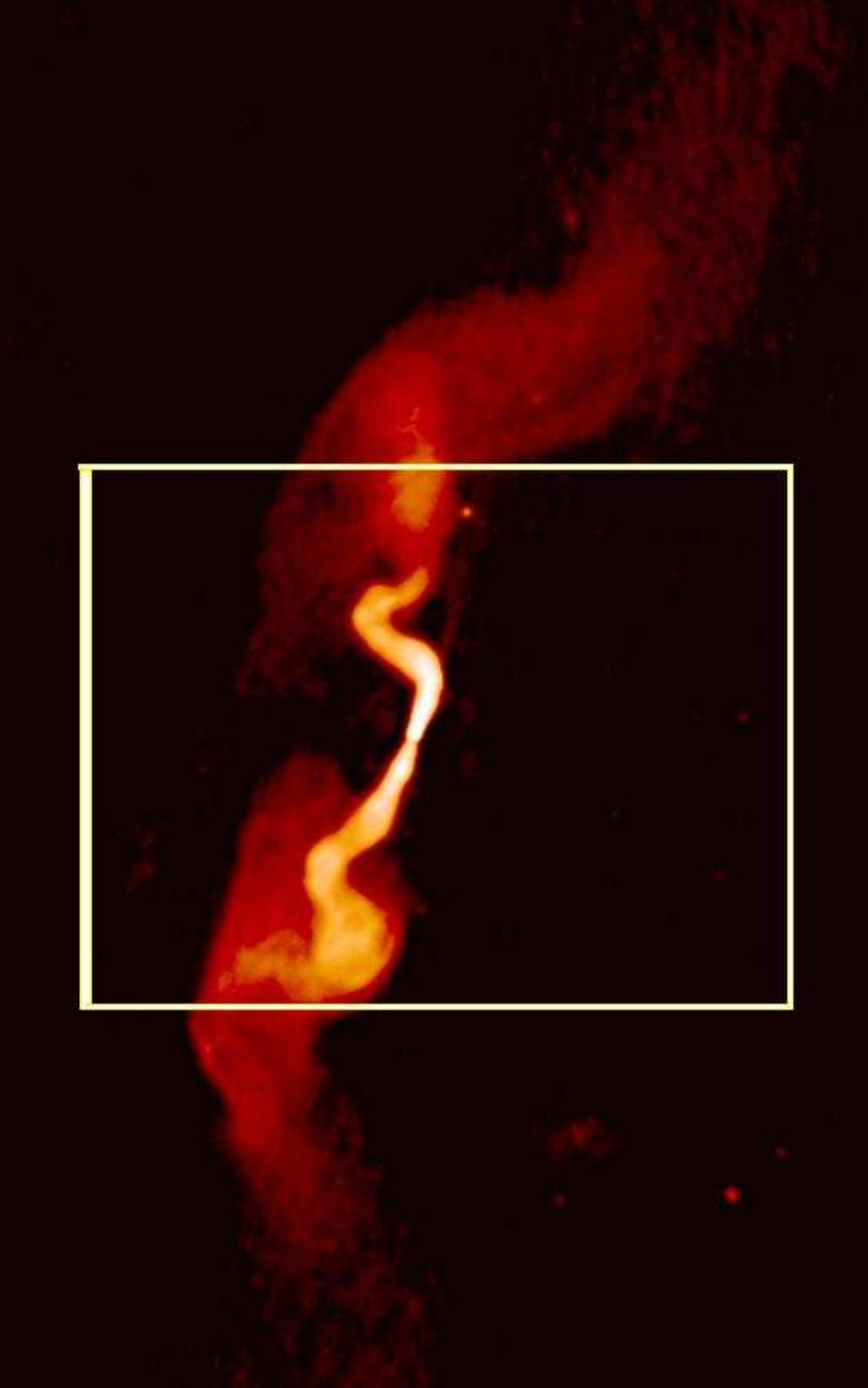}
  \end {center}
\caption {
The large scale structure of 3C31
at 
the frequency of
1.4 GHz. The north-south  field
covers 300~$kpc$ and the small box here modeled
covers  120~$kpc$.
This image was generated with data from telescopes of the National Radio
Astronomy Observatory, a National Science Foundation Facility, managed by
Associated Universities. Inc    .}
  \label{3C31_box}
    \end{figure}

\begin{figure}
  \begin{center}
\includegraphics[width=10cm]{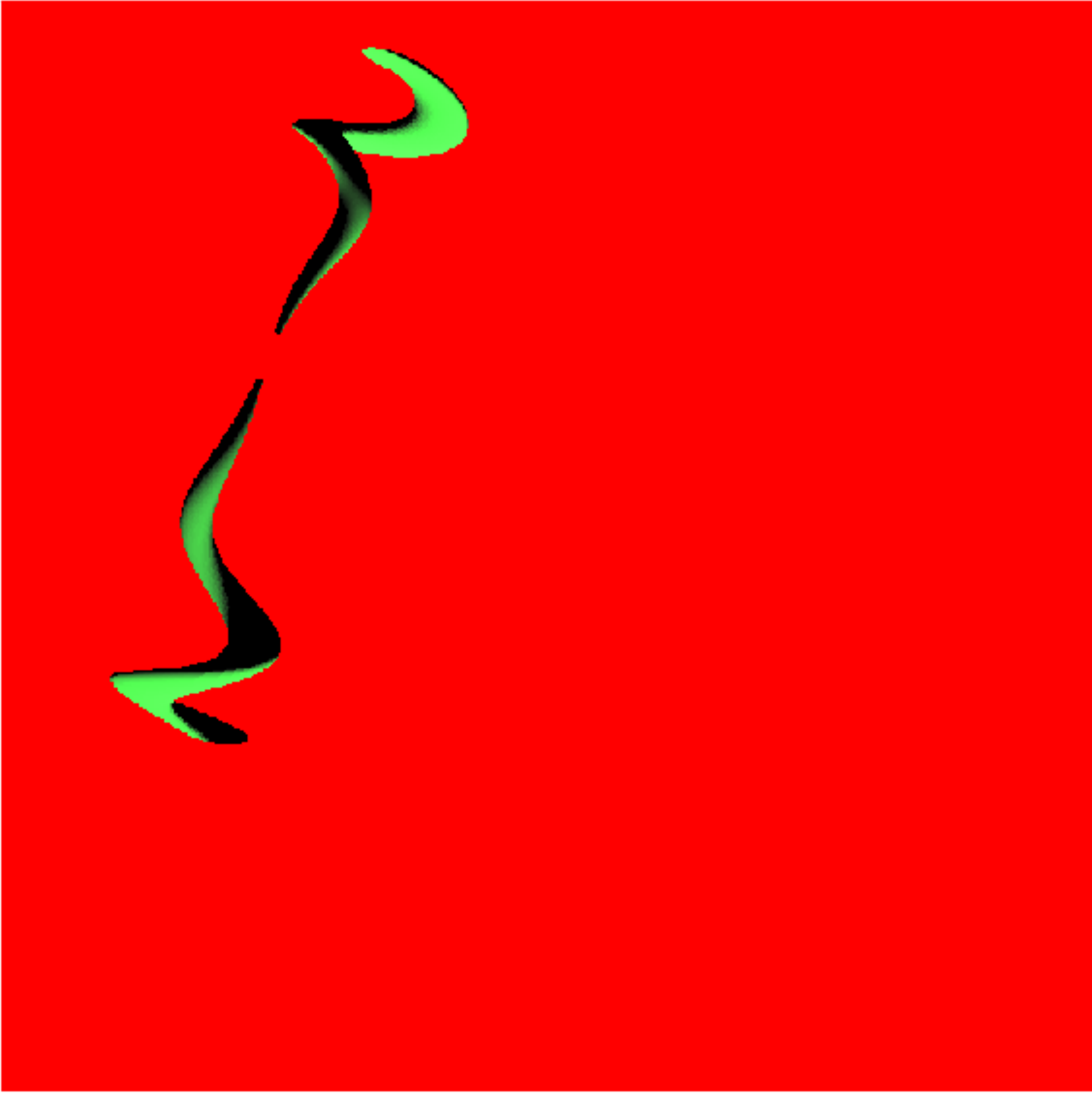}
  \end {center}
\caption {
Continuous  three-dimensional surface of 3C31:
 the three Eulerian angles 
 characterizing the point of view are $ \Phi $=180   $^{\circ }$
 , $ \Theta $=170   $^{\circ }$
 and  $ \Psi $=195  $^{\circ }$.
 The precession is characterized by the angle
  $ \Psi_{prec} $=  10   $^{\circ }$
 and by the angular velocity
 $ \Omega_{prec} $= 54.00 [$^{\circ}/10^7 \mathrm{year}$].
 The three Eulerian angles  which characterize  the jet are
      $ \Phi_{j} $=  0   $^{\circ }$
 ,    $ \Psi_{j} $=  0   $^{\circ }$
 and  $ \Theta_{j} $= 90   $^{\circ }$.
 The angle of rotation of the galaxy is
 $ \alpha_{G} $=  0   $^{\circ }$.
 The physical parameters characterizing the jet motion
 are : $t_7$= 10.00~$10^7 \mathrm{year}$,
       $d_1$=  0.2~${\mathrm{kpc}}$ ,
       $\beta_0$ = 0.05               ,
       $\alpha_0$= 10.64   $^{\circ }$.
The length of the jet is  59.65 ${\mathrm{kpc}}$  and 
the velocity of the galaxy $v_z$= 292 ${\mathrm{km/s}}$ .
          }%
    \label{3C31_surface}
    \end{figure}

\begin{figure}
  \begin{center}
\includegraphics[width=10cm]{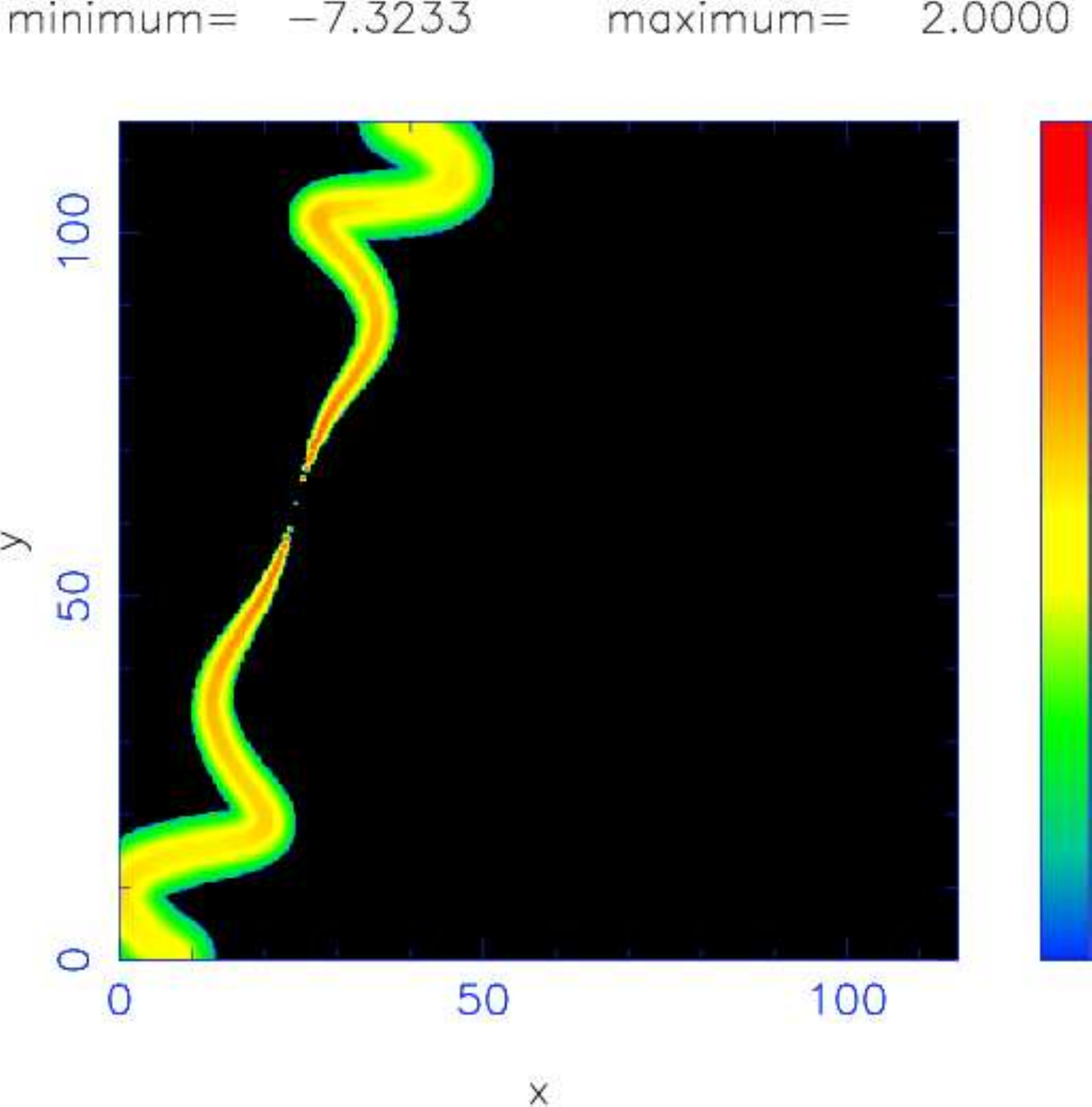}
  \end {center}
\caption {
Theoretical 2D map of the decimal logarithm of the
 intensity of emission
representing the continuous three-dimensional trajectory of 3C31;
parameters as in Figure~\ref{3C31_surface} .
The integral operation is performed on a cubic  grid 
of $500^3$ pixels.
}
  \label{3C31_log}
    \end{figure}

\begin{figure}
  \begin{center}
\includegraphics[width=10cm]{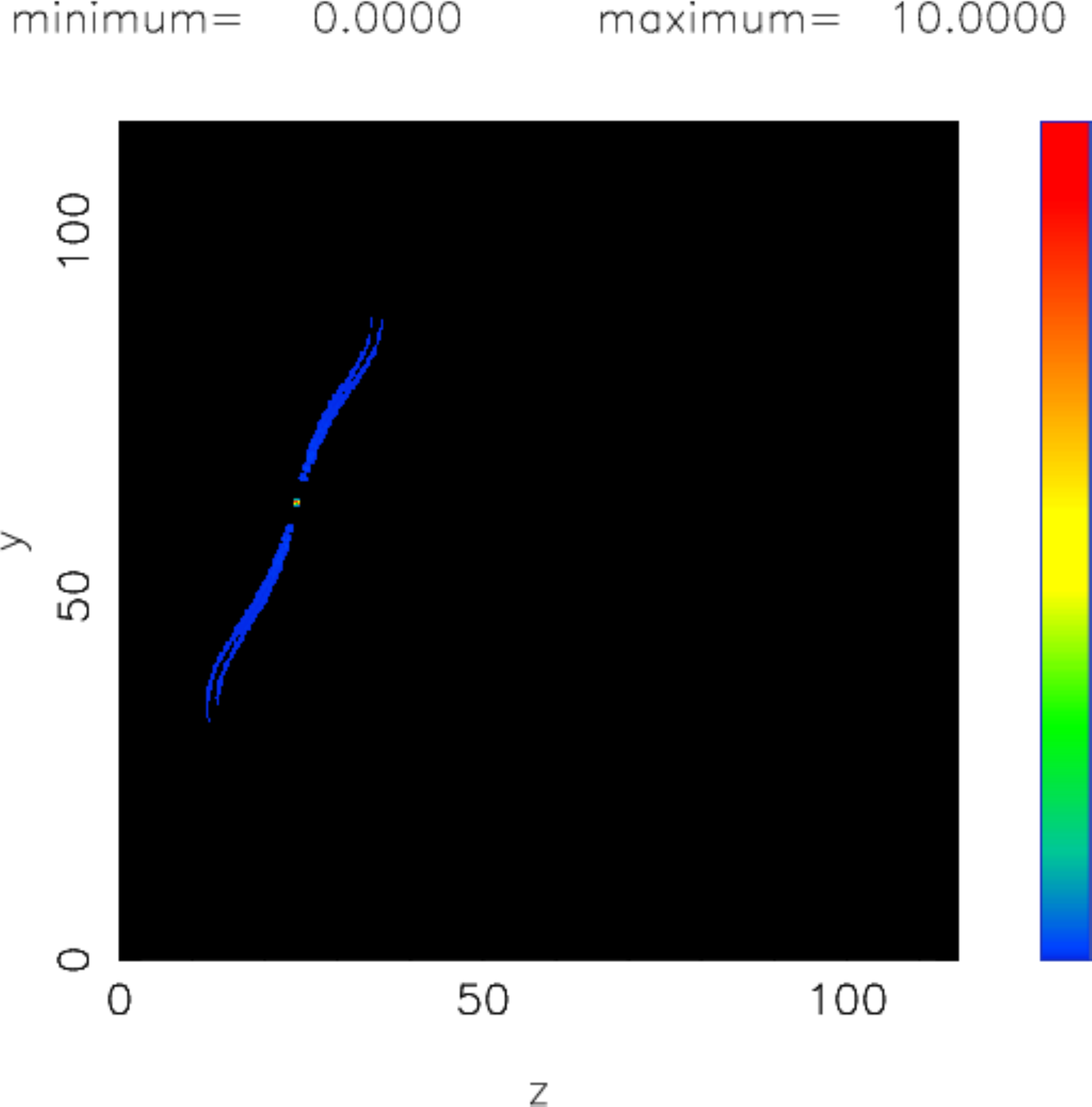}
  \end {center}
\caption {
Sobel filter of the intensity  of the   central part of  3C31.
The gradient magnitude is computed according to formula~(\ref{gradient}),
parameters as in Figure~\ref{3C31_surface} .
}
  \label{3C31_sobel}
    \end{figure}

\begin{figure}
  \begin{center}
\includegraphics[width=10cm]{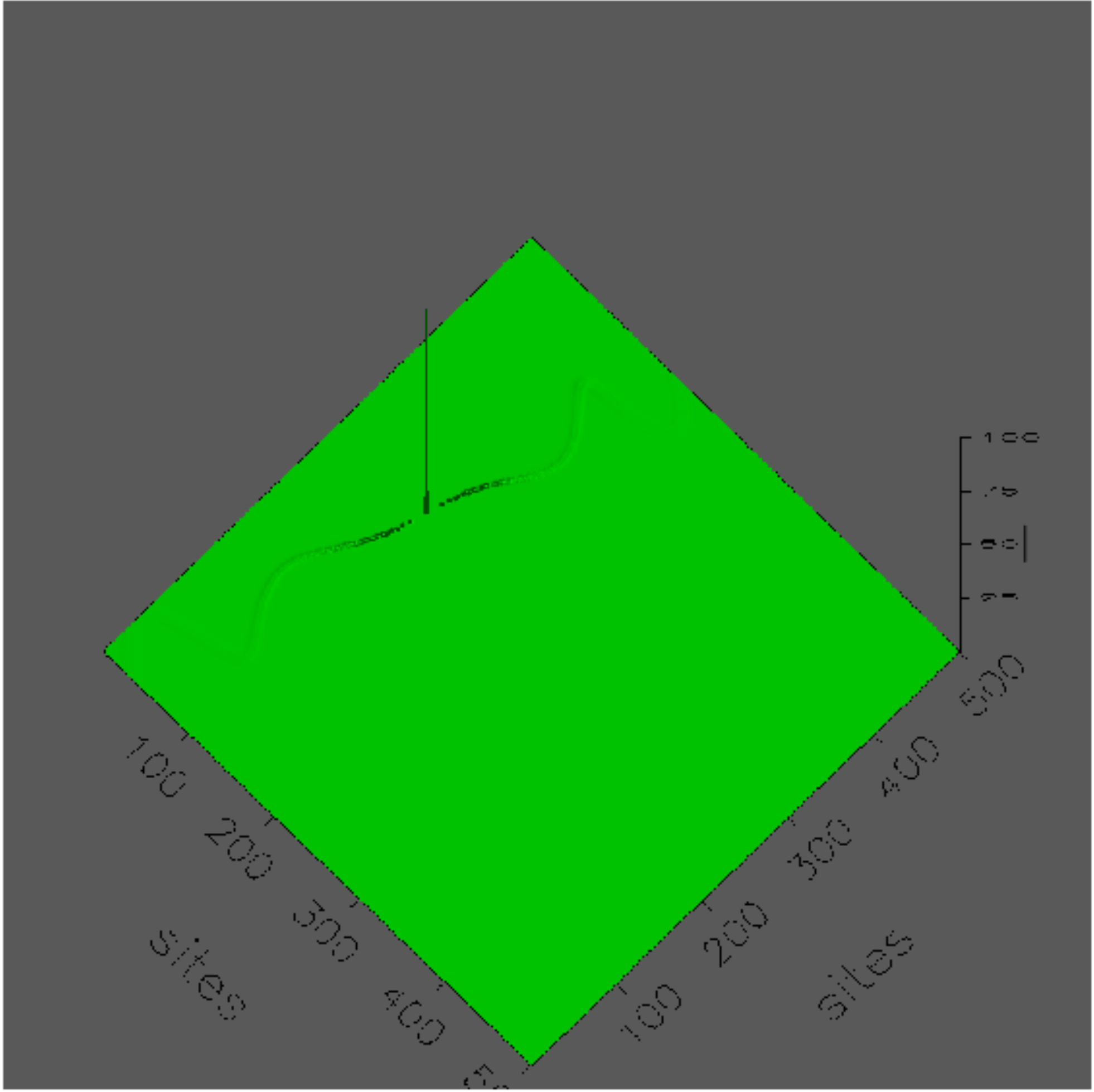}
  \end {center}
\caption {
Intensity of 3C31  represented
through a surface , parameters as  
in Figure~\ref{3C31_surface} and Figure~\ref{3C31_log}.
The  observer sees  the surface
at   an angle of 70$^{\circ }$.  
}
  \label{3C31_euler_surf}
    \end{figure}

\begin{figure}
  \begin{center}
\includegraphics[width=10cm]{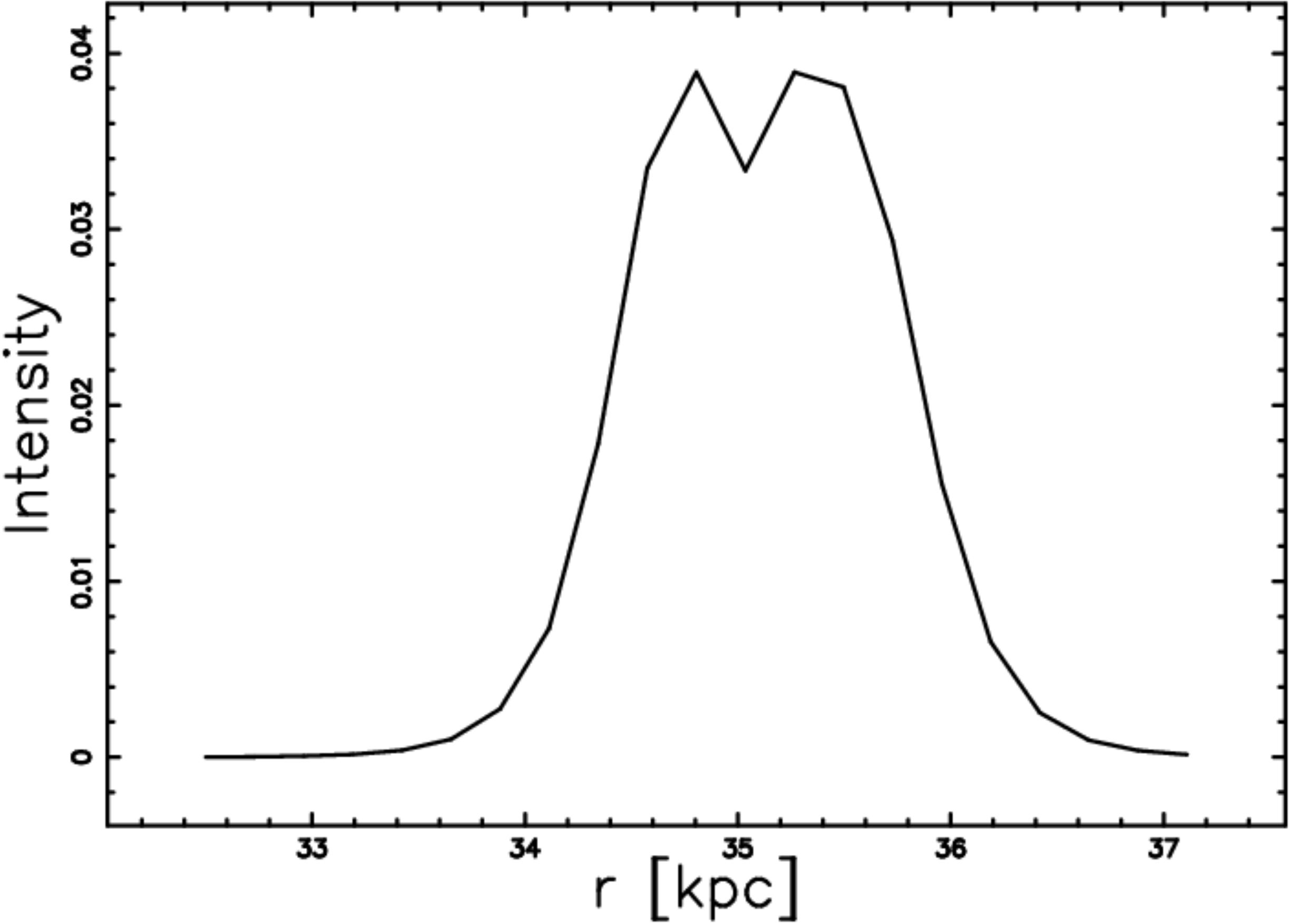}
  \end {center}
\caption {
Intensity of 3C31  represented
through a 1D cut along the x-direction , parameters as  
in Figure~\ref{3C31_surface} and Figure~\ref{3C31_log}.
}
  \label{3C31_cut}
    \end{figure}

Another radio-galaxy that can be here simulated is 
3C31 ; this radio-galaxy has been chosen because it does not
present  knots, see radio observations in
Figure~\ref{3C31_box}.
This radio-galaxy has already been simulated in the first 
straight 12 $kpc$ , and a comparison
between the model and the observed brightness distribution 
is  reported 
in Figure~6 of~\citet{Laing2002} 
and in  Figure~3 of~\citet{Laing_2005}~.
The first 60~$kpc$ are simulated  in order  to cover the wiggles.
The 3D surface that represents the trajectory is presented
in  Figure~\ref{3C31_surface},
the simulated  intensity distribution 
is reported in Figure~\ref{3C31_log}  , 
Figure~\ref{3C31_sobel}
reports the gradient magnitude
computed according to the Sobel filter,
Figure~\ref{3C31_euler_surf} reports the intensity 
represented through a surface rendering
that visualizes the "valley on the top",
and  Figure~\ref{3C31_cut} reports the intensity 
represented through a 1D cut.

\subsection{Spectral index}

The collision of an
electron of energy $E$  with a cloud
produces an average gain in energy ($\Delta E$)
proportional to the second order in $u/c$ ,
$u$ is cloud velocity and $c$ is light velocity,
\begin {equation}
\langle \frac {\Delta E}  {E } \rangle
=
\frac {8}{3} ( \frac {u} {c} )^2  \quad.
\end  {equation}
This  process is named  Fermi~II after \citet{Fermi49}.

On introducing  an average length  of collision
, $\lambda$ ,  the
formula becomes:
\begin  {equation}
\frac {d  E}  {dt }
=
\frac {E }  {\tau }   \quad,
\end {equation}
where
\begin {equation}
\tau  = \frac {4} {3 } ( \frac {u^2} {c^2 }) (\frac {c } {\lambda })
\quad.
\label {tau}
\end   {equation}
It should be remembered  that  with a mean free path between clouds
$\lambda$,  the average time between collisions
with clouds , see \citet{lang} ,  is
\begin{equation}
2\frac{\lambda}{c}
\quad  .
\label{twice}
\end{equation}

The probability , $P(t)$ , of  the particle remaining  in the reservoir
for a period greater than $t$ is,
\begin {equation}
P(t) = e ^{- \frac {t} {T}}
\quad  ,
\end   {equation}

where $T$ is the time of escape  from the considered region.
The hypothesis that
 energy is continuously
injected in the form of relativistic particles with energy
$E_0$ at the rate R, produces ( according to 
\citet{Burn75})
the following probability density , $N$ ,
which is  a  function of   the energy:
\begin {equation}
N (E)  =
\frac {R \tau} {E_0}
\left (\frac {E} {E_0} \right )^{-\gamma}
\quad  ,
\label {eq:ne}
\end   {equation}
with
\begin {equation}
\gamma = 1 + \frac {\tau} {T }
\quad,
\end   {equation}
and $\tau$ is defined in equation~(\ref{tau})~.
Equation~(\ref{eq:ne}) can be written as
\begin {equation}
N (E)  =K \;  E ^{-\gamma}
\quad  ,
\label {eq:neK}
\end   {equation}
where $K = \frac {R \tau} {{E_0}^{-\gamma +1}}$~.
A power law spectrum in the electron  energy
has now been obtained.
A power law distribution of relativistic electrons ,see \citet{lang} ,
corresponds to a power law frequency spectrum 
proportional to $\nu ^{\alpha_s}$ , where the spectral index $\alpha_s$
is 
\begin{equation}
\alpha_s =- \frac{\gamma -1 }{2} =- \frac{\tau}{2T}
\quad .
\label{indicealfa}
\end{equation}

\begin{figure}
  \begin{center}
\includegraphics[width=10cm]{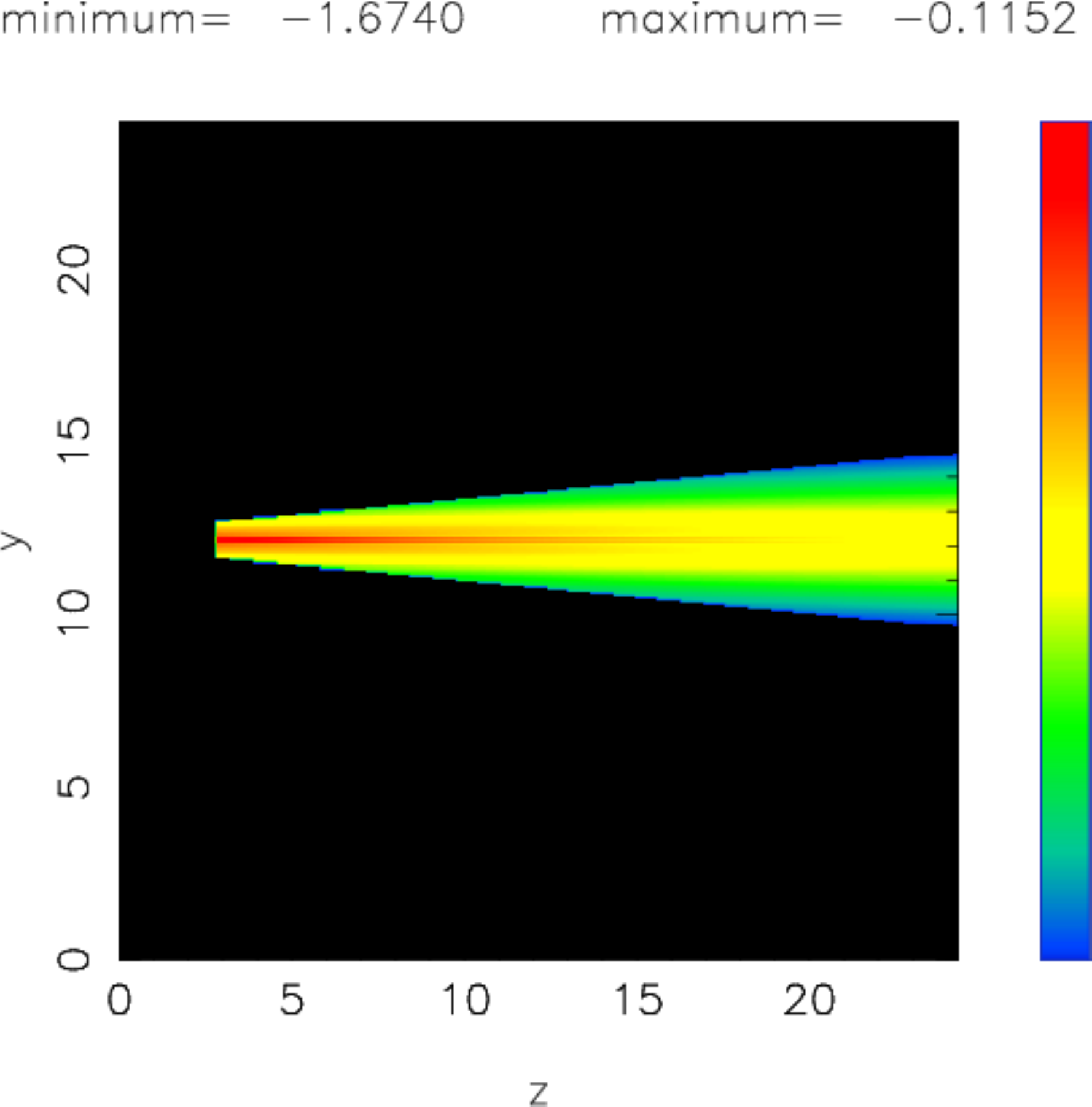}
  \end {center}
\caption {
Cut of the first spectral index through a plane 
crossing the center of the jet, $b$=14.
The length of the jet is $24kpc$ , the same as 3C273.
The 2D    grid is made  
of $400^2$ pixels.

}
  \label{3C273_alfa_cut}
    \end{figure}

\begin{figure}
  \begin{center}
\includegraphics[width=10cm]{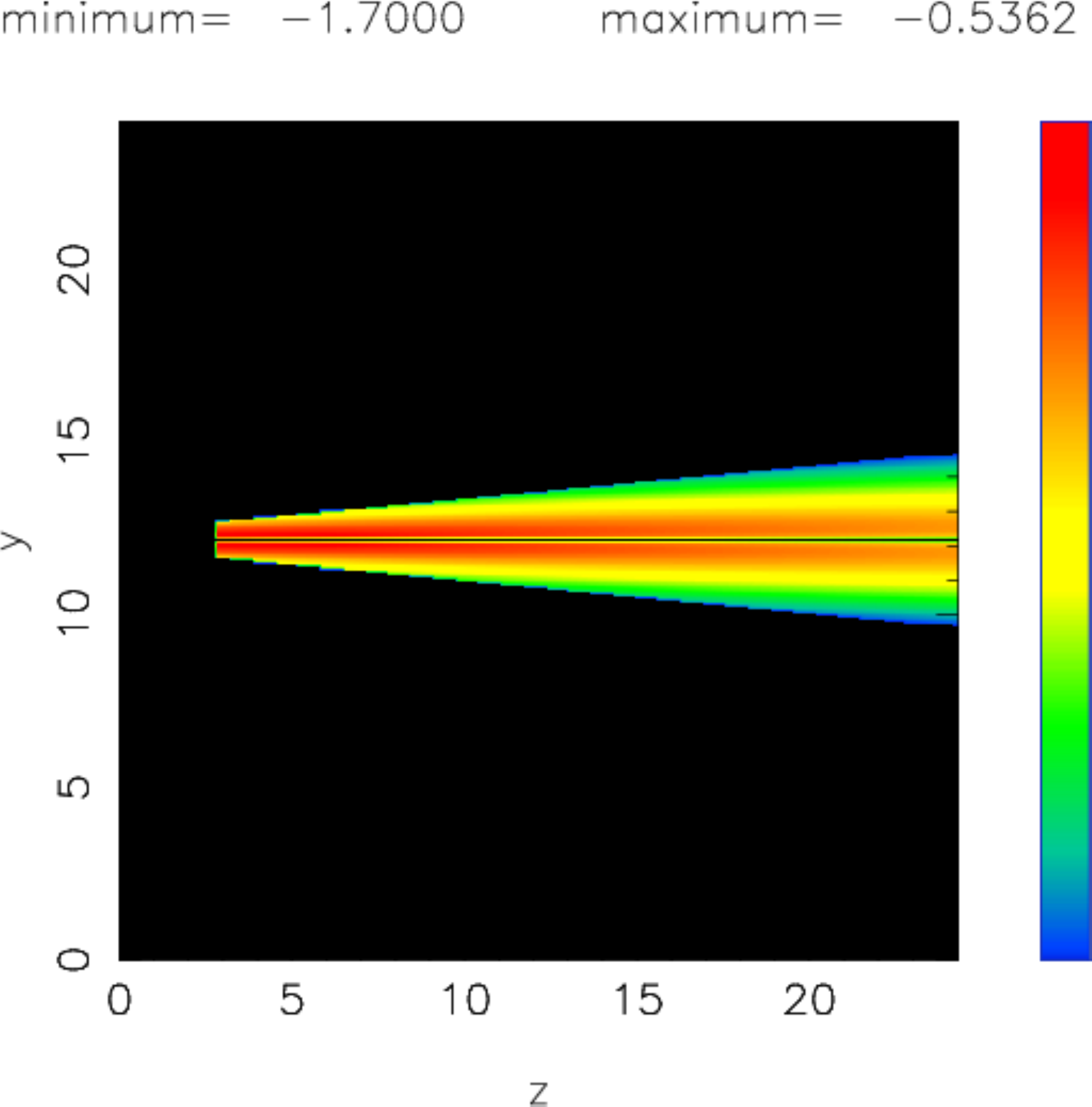}
  \end {center}
\caption {
Spectral index of 3C373 as comes  from the integral operation
over two frequencies.
The integral operation is performed on cubic  grid 
of $400^3$ pixels.
}
  \label{3C273_alfa_global}
    \end{figure}

 $T$ is now assumed to be  constant and 
$\tau$    scales  with the 
power released in turbulent kinetic energy, $\epsilon$ , 
as 
\begin{equation}
\tau \approx \frac{1}{[\epsilon(r,z)]^{1/b}}
\quad  ,
\end{equation}
where $b$ is a parameter that connects theory with observations.
Concerning the observations  of the spectral index 
along the jet we concentrate on 3C273 , 
see  X-observations by~\citet{Jester2006} and far-ultraviolet
imaging  by~\citet{Jester2007}.
 Table~3 in X-Ray
by~\citet{Jester2006} was selected as an example 
to simulate : the spectral index is 0.83 
at knot $B1$ (internal region)  and 1.27 at knot $H2$ (external region).
Therefore as suggested by the   theoretical formula~(\ref{indicealfa}) 
the observed spectral index increases in absolute value going from the inner
regions to the outer regions.
The theoretical spectral index was simulated in the following way 
\begin{enumerate}
\item A spectral index is generated according to 
      formula~(\ref{indicealfa})  in each point of the 3D jet.
      A typical 2D cut is reported in Figure~\ref{3C273_alfa_cut}.
\item The intensity at the frequency $\nu$, $I_{\nu}$ , is computed 
      in each point of the jet according to 
      equation~(\ref{e_astro}).
\item The intensity at a frequency one decade bigger , $\nu \times 10$,
      is evaluated according to the scaling ,
      $I_{\nu \times 10} = I_{\nu} (10)^{\alpha_s}$~.
\item The integral operation is performed on a cubic grid 
      of $pixels^3$ points. The spectral index that 
      comes  from the two integral operations at two 
      different frequencies  is  computed ,
      see Figure~\ref{3C273_alfa_global}.
      The parameter $b$ is chosen in order to fit the observations.
\end{enumerate}

In this simulation , see Figure~\ref{3C273_alfa_global} ,
the spectral 
index increases in absolute value both in going from 
the internal regions to the external regions 
following the axial direction and  in going from the center 
of the jet to the boundary along  the perpendicular direction.
This effect has been observationally mapped in the case 
of 3C296 , see for example Figure 4b in~\citet{Laing_2006}~.

\begin{figure}
  \begin{center}
\includegraphics[width=10cm]{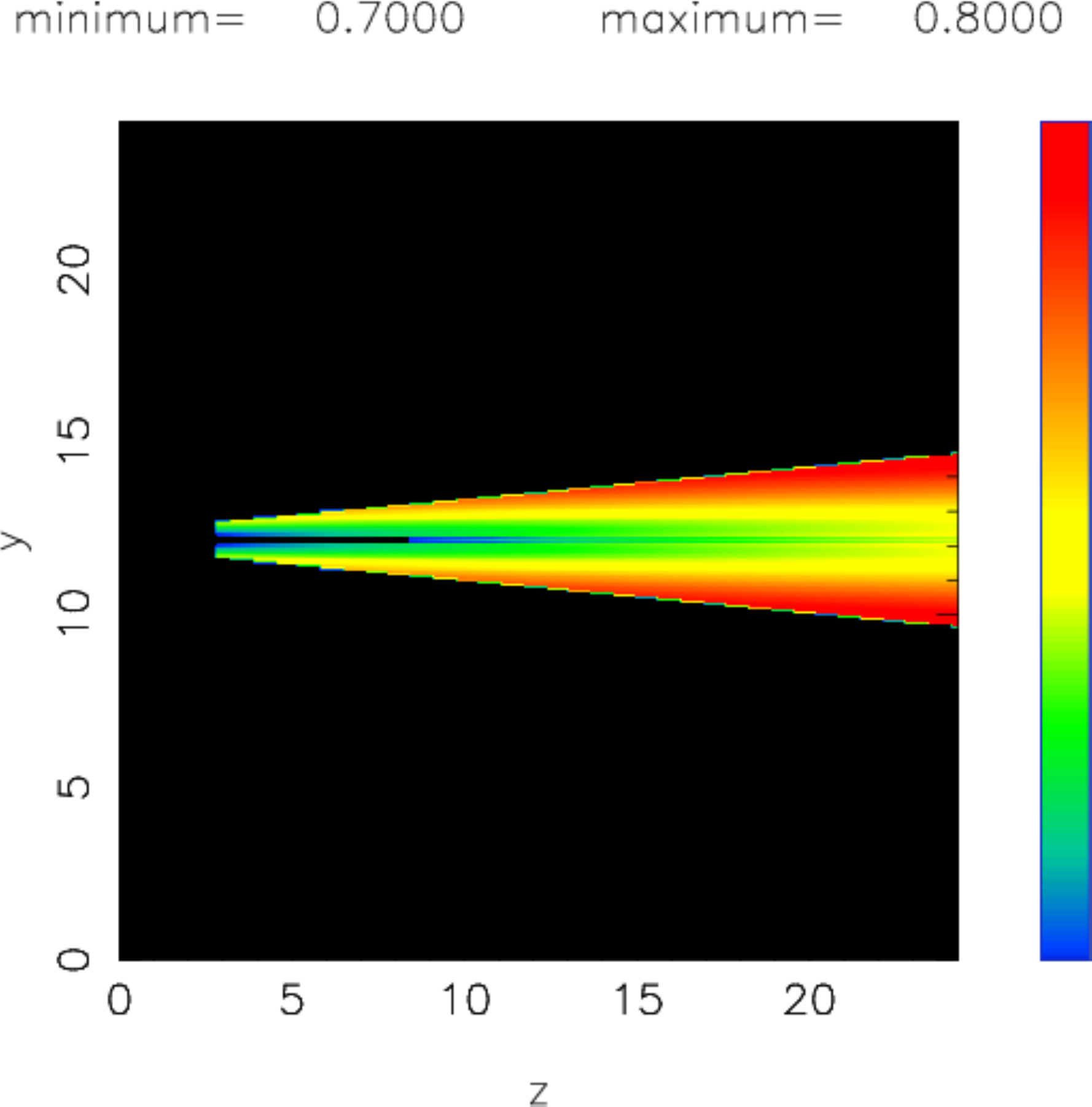}
  \end {center}
\caption {
Degree  of linear polarization  of 3C373;
parameters as in Figure~\ref{3C273_alfa_global}~.
}
  \label{3C273_polariz}
    \end{figure}

Under the hypothesis of ultra-relativistic electrons , 
a power law $-\gamma$  in the  energy distribution 
of the electrons  and a uniform magnetic field, 
the degree of linear polarization is,  
see equation~(1.176) in \citet{lang} ,
\begin{equation}
\prod  = \frac{\gamma +1} {\gamma +7/3} 
\quad .
\end{equation}
Figure~\ref{3C273_polariz} maps the degree  of linear 
polarization  that comes  from the 2D map of the 
spectral index after a conversion of the 
index  $\gamma$ in $\alpha_s$ ,
\begin{equation}
\prod  = \frac{3(1-\alpha_s)} {3 \alpha_s - 5 } 
\quad .
\end{equation}

\section{Summary}

{ \bf Theoretical Emissivity~} The power released in the turbulence is computed for three type of
fluids. The maximum input in turbulence is realized when $ r=0.69
{b_{1/2}}$ , in the case that  turbulent fluids are considered. 

{\bf Theoretical Intensity Profiles~} The
profiles in intensity of synchrotron radiation perpendicular to
the line of sight are analytically or numerically solved when the
simplest configuration is realized : straight jets oriented in the
direction perpendicular to the observer. The maximum of
synchrotron radiation 
for turbulent fluids 
is at $ r \approx 0.49 {b_{1/2}}$. 
This
means that  a depression at the center of the
jet is expected. 
The three types of synchrotron radiation here analyzed are
then compared with the observed X-component  of knot D-E in M87
and with  knot A in  3C273 at
the frequency of
 5GHz. The confrontation with the
data from astronomical observations should be carefully done
because the data are often smoothed and this operation may change
the structure of the central depression. 

{\bf Theoretical Radio maps} More complex is the set
up of an algorithm that implements the radiative transfer equation
for a randomly oriented jet. In order to build such an image, the
centerline trajectory should be made discrete and the distance
between the point of a 3D grid and the nearest point of  the
trajectory computed. This algorithm  computes the
theoretical map in intensity of synchrotron radiation for complex
morphological cases such as NGC4061.
The depression at the center of the jet
, Figure~\ref{NGC4061_euler_surf},
 can be considered an 
alternative to the cosmic double helix , see
\citet{Lobanov2001,Lobanov2005}.

{\bf Theoretical Spectral Index} The connection between power released in turbulence 
and acceleration of electrons has been attached adopting 
a  Fermi mechanism 
with a variable velocity  of  acceleration.
This mechanism, after the introduction of a parameter that
connects theory and observations allows us to build 2D maps
of the theoretical spectral index of synchrotron emission,
see Figure~\ref{3C273_alfa_global}.
The simulation predicts that the spectral index should 
decrease  ( it increases in absolute value)
along the centerline and the transversal directions.
As a consequence  the 2D map of the degree of linear 
polarization is easily built.

{\bf Theoretical Sobel Filter} The new technique  of the Sobel 
Filter can  detect details in the map of radio-galaxies.
The phenomena of the "valley on the top" can be revealed 
due to the   gradient's    
change of sign
   in going 
from the center to the external regions in a radial direction,
see Figure~\ref{blob_turb_3C273_sobel} 
and Figure~\ref{NGC4061_sobel_4}~.
The presence of arcs in the Sobel-filtered images , see Figure~26b
in \citet{Laing_2006} , can be explained by  the fact 
that the integral operation that leads to the intensity of radio-galaxies
is made in presence of curved zones of emission.
\\
\\
I express my gratitude to  Miguel Onorato   for  useful help   
during the   set up  of Figure~2.


\label{lastpage}
\end{document}